\documentclass[manuscript]{aastex}
\usepackage{natbib, aas_macros}
\usepackage[dvips]{color} 
\slugcomment{Not to appear in Nonlearned J., 45.}

\shorttitle{Observations of the Near- to Mid-Infrared Unidentified Emission Bands in the Interstellar Medium of the Large Magellanic Cloud}
\shortauthors{Mori I., et al.}

\begin{document}

\title{Observations of the Near- to Mid-Infrared Unidentified Emission Bands in the Interstellar Medium of the Large Magellanic Cloud}

\author{Tamami I. Mori\altaffilmark{1}, Itsuki Sakon\altaffilmark{1} and Takashi Onaka\altaffilmark{1}}
\affil{Department of Astronomy, Graduate School of Science, The University of Tokyo}
\email{[morii;isakon;onaka]@astron.s.u-tokyo.ac.jp}

\author{Hidehiro Kaneda\altaffilmark{2}}
\affil{Graduate School of Science, Nagoya University}

\author{Hideki Umehata\altaffilmark{3}}
\affil{Institute of Astronomy, Graduate School of Science, The University of Tokyo}

\and

\author{Ryou Ohsawa\altaffilmark{1}}
\affil{Department of Astronomy, Graduate School of Science, The University of Tokyo}

\altaffiltext{1}{Department of Astronomy, Graduate School of Science, The University of Tokyo, 7-3-1 Hongo, Bunkyo-ku, Tokyo 113-0033, Japan}
\altaffiltext{2}{Graduate School of Science, Nagoya University, Chikusa-ku, Nagoya 464-8602, Japan}
\altaffiltext{3}{Institute of Astronomy, Graduate School of Science, The University of Tokyo, 2-21-1 Osawa, Mitaka, Tokyo 181-0015, Japan}

\begin{abstract}
We present the results of near- to mid-infrared slit spectroscopic observations (2.55--13.4\,$\mu$m) of the diffuse emission toward nine positions in the Large Magellanic Cloud with the Infrared Camera (IRC) on board {\it AKARI}. 
The target positions are selected to cover a wide range of the intensity of the incident radiation field. 
The unidentified infrared bands at 3.3, 6.2, 7.7, 8.6 and 11.3\,$\mu$m are detected toward all the targets, and ionized gas signatures: hydrogen recombination lines and ionic forbidden lines toward three of them. 
We classify the targets into two groups: those without the ionized gas signatures (Group A) and those with the ionized signatures (Group B). 
Group A includes molecular clouds and photo-dissociation regions, whereas Group B consists of \ion{H}{2} regions. 
In Group A, the band ratios of  $I_{\rm 3.3\,\mu m}$/$I_{\rm 11.3\,\mu m}$, $I_{\rm 6.2\,\mu m}$/$I_{\rm 11.3\,\mu m}$,  $I_{\rm 7.7\,\mu m}$/$I_{\rm 11.3\,\mu m}$ and $I_{\rm 8.6\,\mu m}$/$I_{\rm 11.3\,\mu m}$ show positive correlation with the {\it IRAS} and {\it AKARI} colors, but those of Group B do not follow the correlation. 
We discuss the results in terms of the polycyclic aromatic hydrocarbon (PAH) model and attribute the difference to the destruction of small PAHs and an increase in the recombination due to the high electron density in Group B. 
In the present study, the 3.3\,$\mu$m band provides crucial information on the size distribution and/or the excitation conditions of PAHs and plays a key role in the distinction of Group A from B. 
The results suggest the possibility of the diagram of $I_{\rm 3.3\,\mu m}$/$I_{\rm 11.3\,\mu m}$ v.s. $I_{\rm 7.7\,\mu m}$/$I_{\rm 11.3\,\mu m}$ as an efficient diagnostic tool to infer the physical conditions of the interstellar medium.

\end{abstract}

\section{Introduction}
Since the discovery of the 11.3\,$\mu$m band in planetary nebulae in 1973 \cite[]{gill73}, 
space infrared missions, and air-borne and ground-based infrared observations have shown that the major unidentified infrared (UIR) bands appear at 3.3, 6.2, 7.7, 8.6, 11.3, 12.6 and 16.4\,$\mu$m together with some faint features. 
The UIR bands have been observed in various astrophysical environments, including photo-dissociation regions (PDRs), reflection nebulae, planetary nebulae \citep[e.g.,][]{peet02}, the diffuse interstellar medium (ISM) \citep[e.g.,][]{onak96,matt96}, nearby galaxies of various types \citep[e.g.,][]{dale06,kane08,smit07}, and distant galaxies \citep[e.g.,][]{lutz05,saji07}. 
The carriers of the UIR bands are generally thought to be polycyclic aromatic hydrocarbons (PAHs) or PAH-containing carbonaceous compounds \citep[e.g.,][]{saka84,pule89,papo89,alla89}. 
PAHs are excited by absorbing a single UV photon and emit a number of IR photons corresponding to vibration modes of C-C and C-H bonds. 
The 3.3\,$\mu$m band is assigned to C-H stretching modes, 
the 6.2\,$\mu$m band to C-C stretching modes, 
the 7.7\,$\mu$m band to blending of several C-C stretching modes and C-H in-plane bending modes, 
the 8.6\,$\mu$m band to C-H in-plane bending modes, 
the 11.3\,$\mu$m band to solo C-H out-of-plane bending modes, and
the 12.6\,$\mu$m band to trio C-H out-of-plane bending modes, 
respectively \citep{alla89}. 

Recent laboratory experiments and quantum chemical calculations suggest that the properties of the UIR bands (e.g., shapes, center wavelengths, interband ratios, etc.) reflect the chemical and physical properties of PAHs (e.g., molecular structure, size distribution, ionization state, temperature, etc.), which may be altered in interstellar and circumstellar environments \cite[]{tiel08}. 
Therefore the UIR bands have a great potential to be used as efficient diagnostic tools to infer the physical condition of the ISM even in remote galaxies. 

Observations of the diffuse Galactic radiation and normal galaxies have shown very little variation in the  mid-infrared (MIR) UIR band spectra (6--12\,$\mu$m) until recently \citep{chan01,luna03,sako04}, whereas small variations in the MIR UIR bands have been reported between the disk and halo regions or the arm and interarm regions of galaxies \citep{irwi06,sako07}. 
Latest {\it Spitzer} and {\it AKARI} observations clearly show distinct variations in the MIR UIR spectra in particular galaxies \citep{kane07,smit07,gall08} for the first time. 
However, the diagnostic for the physical conditions of the ISM as well as the chemical and physical evolution of PAHs in galaxies by means of the UIR bands is not yet fully explored. 

In this paper, we present the results of near-infrared (NIR) to MIR spectroscopic observations of the ISM with different radiation conditions in the Large Magellanic Cloud (LMC) with the infrared camera (IRC) onboard {\it AKARI} \citep[]{mura07,onak07}. 
The LMC is a nearby irregular galaxy and is located at the distance of 50\,kpc from the Milky Way \citep{feas99,kewo06}. 
In addition to its proximity, the almost face-on orientation \citep[$i\sim$35$^\circ$;][]{vand01,olse02,niko04}
provides us with a unique opportunity to investigate regions with different physical conditions without confusion because of the spatial resolution of $\sim5\arcsec$ in the MIR of {\it AKARI}/IRC. 
The IRC spectroscopy has a unique characteristic that it can obtain a spectrum from 2.5 to 13\,$\mu$m  simultaneously  with the same slit \citep{ohya07}. 
This has an advantage for the spectroscopy of extended objects over other space instruments since the Short Wavelength Spectrometer (SWS) onboard the {\it Infrared Space Observatory} ({\it ISO}) had different diaphragms from the NIR to MIR \citep{degr96} and the Infrared Spectrograph (IRS) on {\it Spitzer} lacks a channel in the NIR \citep{houc04}. 
\cite{verm02} report the MIR UIR band ratios of \ion{H}{2} regions of the LMC based on observations with the ISOPHOT/PHT-S instrument on board  {\it ISO}, but do not include the 3.3\,$\mu$m band in their analysis because of the low signal-to-noise ratio (S/N) in the short wavelength channel. 
The 3.3\,$\mu$m band, in fact, is most sensitive to the smallest PAHs \citep[e.g.,][]{schu93} and its relative intensity to the MIR UIR bands provides us with significant information on the average temperature of PAHs, which depends on the size distribution and the excitation condition of PAHs. 
In this paper, we investigate variations in the relative intensity of the UIR bands in the NIR to MIR of the diffuse radiation from regions with different radiation field conditions in the LMC and discuss them in relation to the physical properties of PAHs. 

In \S 2, the observation and the data reduction are described together with the selection of the target positions. 
The obtained spectra are presented in \S 3.  
In \S 4, the observed variations in the UIR band ratios in different radiation field conditions are investigated in terms of the PAH model.  
Diagnostic of the physical conditions of the observed regions is also discussed based on the UIR band ratios. A summary and conclusions are given in \S 5. 

\section{Observations and Data Reduction}
\subsection{Observations}
The present study employs the datasets of eight pointed observations (observation IDs: 1400330, 1400346, 1400318, 1402426, 1400324, 1402422, 1400334 and 1400320) collected as part of the {\it AKARI} mission program  "ISM in our Galaxy and Nearby Galaxies" \citep[ISMGN;][]{kane09}. 
All of the observations were performed with the slit spectroscopic mode with the choice of the grism for the NIR disperser \citep[AOT04 b:Ns;][]{ohya07}. 
The NIR spectrum was taken with the grism, NG (2.5--5.0\,$\mu$m, $\lambda/\Delta\lambda\sim100$), in the NIR channel and the MIR spectra were taken with two grisms, SG1 (4.6--9.2\,$\mu$m, $\lambda/\Delta\lambda\sim50$) and SG2 (7.2--13.4\,$\mu$m, $\lambda/\Delta\lambda\sim50$), in the MIR-S channel. 
The NIR and MIR spectra were obtained simultaneously by means of the beam splitter, which enabled us to get a continuous NIR to MIR spectrum from 2.5 to 13.4\,$\mu$m of the same slit area of 1$\arcmin$ length by 5\,$\arcsec$ width \citep[]{ohya07}. 
 The observation parameters are summarized in Table \ref{tbl_obs}. 
 The accurate slit position is determined from the 3.2\,$\mu$m image (N3), which is taken during each pointed observation (see \S \ref{red_nir}), by referring to the positions of point sources in the 2MASS catalog.

\subsection{Target Selection}
The targets are selected by taking account of the CO mapping data \citep{mizu01} and the {\it IRAS} colors of $I_{\rm 25\,\mu m}$/$I_{\rm 12\,\mu m}$ and $I_{\rm 60\,\mu m}$/$I_{\rm 100\,\mu m}$, which indicate local star formation activities \citep{boul88,onak07-2}. 
\cite{sako06} have shown that the extremely large {\it IRAS} colors of $I_{\rm 25\,\mu m}$/$I_{\rm 12\,\mu m}$ and $I_{\rm 60\,\mu m}$/$I_{\rm 100\,\mu m}$ 
($\sim$3 and $\sim$0.6, respectively) in the diffuse emission in the LMC can be accounted for by a large contribution from nearby young ($< 30\,{\rm Myr}$) clusters to the incident interstellar radiation field and that the small {\it IRAS} colors of $I_{\rm 25\,\mu m}$/$I_{\rm 12\,\mu m}$ and $I_{\rm 60\,\mu m}$/$I_{\rm 100\,\mu m}$ ($\sim$1 and $\sim$0.3, respectively) are those expected from the heating by the incident radiation field inside quiescent molecular clouds \citep[e.g.,][]{mivi02}. 
According to these results, we select several infrared bright positions with different {\it IRAS} colors as the targets of the present study, where molecular clouds are recognized on the CO maps \citep{mizu01,fuku08}. 
Since the beam size of the {\it IRAS} data is larger than the size of the slit of the {\it AKARI}/IRC, we also derive the {\it AKARI} color of $I_{\rm L24}$/$I_{\rm S11}$ from the dataset of \cite{itay08}, where $I_{\rm L24}$ is the flux density at the L24 (24\,$\mu$m) band and $I_{\rm S11}$ is that at the S11 band (11\,$\mu$m) of the {\it AKARI}/IRC, to obtain the local radiation field conditions. 
The flux density is measured over an aperture of $5\arcsec$ in diameter around the central position of the slit. 
The {\it IRAS} and {\it AKARI} colors of the present targets are summarized in Table \ref{tbl_tgt_imf}. 
The trend of the {\it AKARI} color of $I_{\rm L24}$/$I_{\rm S11}$ is very similar to that of the {\it IRAS} $I_{\rm 25\,\mu m}$/$I_{\rm 12\,\mu m}$ color, whereby we confirm that the selection based on the {\it IRAS} colors is in fact relevant to the purpose of the present study. 
Hence, the target positions cover a wide range of incident radiation field conditions, including molecular clouds, PDRs and \ion{H}{2} regions.

\begin{figure}
\epsscale{0.8}
\plotone{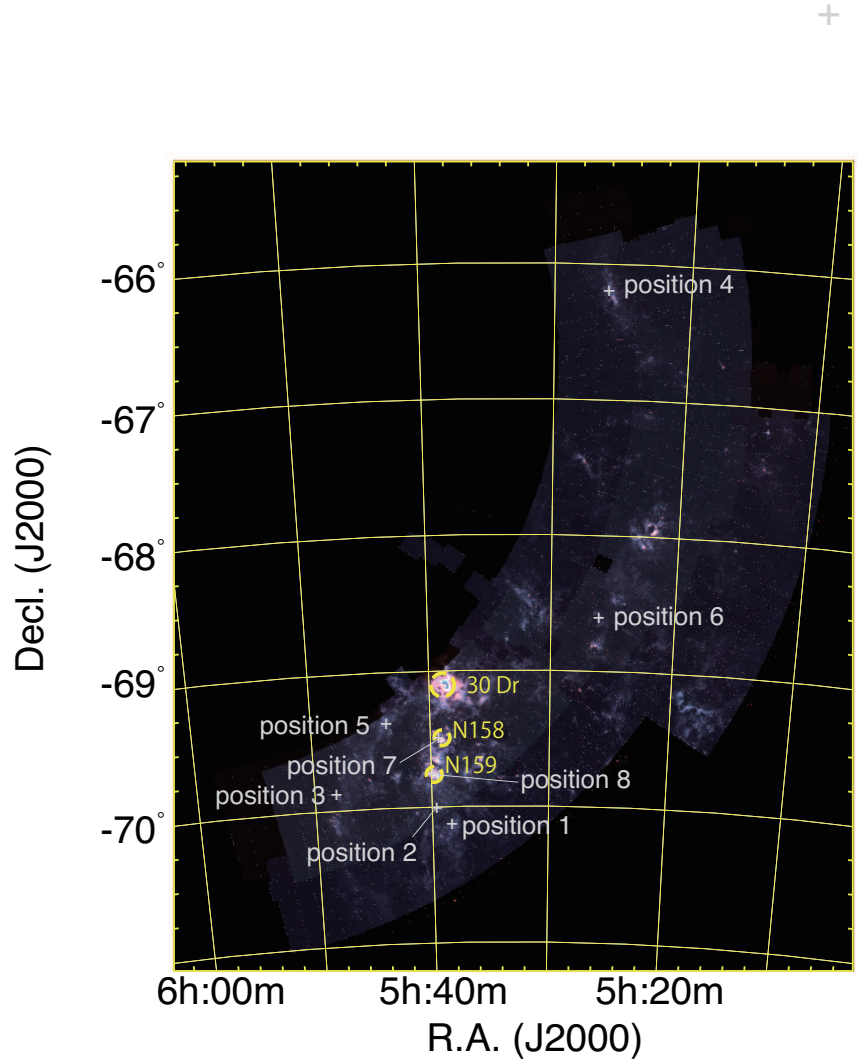}
\caption{The slit positions of each pointed observation are overlain on the artificial 3-color image of the LMC, where the intensity of S7 (7\,$\mu$m) is indicated in blue, S11 (11\,$\mu$m) in green, and L24 (24\,$\mu$m) in red. 
The white crosses indicate the center positions of the slits. 
\label{three_color}
}
\end{figure}

In Figure \ref{three_color}, the slit positions of those datasets are shown over the false color image of the LMC obtained by the {\it AKARI} IRC LMC survey program \cite[]{itay08}. 
The S11 band images of a $10\arcmin\times10\arcmin$ area including each slit position are also shown in Figures \ref{s11}a--h. 
The slit is positioned at regions without apparent point sources in all the positions except for Position 8. 
At Position 8, a bright point-like source is recognized in part of the slit. 
We split the spectrum of Position 8 into 2 parts: the one including the point-like source (Position 8-1) and the other without the point-like source (Position 8-2).  
For the other positions, the spectrum is extracted over the almost entire slit length of 40--50$\arcsec$.

The {\it AKARI} color of  $I_{\rm L24}$/$I_{\rm S11}$ and the {\it IRAS} colors of $I_{\rm 25\,\mu m}$/$I_{\rm 12\,\mu m}$ and $I_{\rm 60\,\mu m}$/$I_{\rm 100\,\mu m}$ at Positions 1, 2, 3, 4, 5 and 6 exhibit only a limited range of 0.5--1.8, 1.0--1.9, and 0.3--0.5, respectively. 
These targets are not associated with SWB 0 or I type star clusters in \cite{bica96}, indicating that they are surrounded by relatively quiescent environments \cite[]{sako06}. 
The {\it AKARI} and {\it IRAS} colors at Positions 7, 8-1 and 8-2 exhibit large values. 
These targets are located in regions associated with OB star clusters as well as Herbig Ae/Be star clusters \citep[N158-O1 and N158-Y1 at Position 7 and N159-Y4 at Position 8;][]{naka05}, confirming that the infrared colors manifest recent star-formation activities in the ISM.

\subsection{Data Reduction}
The present data reduction basically follows the standard toolkit for the IRC spectroscopy. 
However, most of the present targets are faint and require careful data processing. 
Thus, some part of the process is carried out separately from the toolkit with special care. 
Details of the data reduction process are described in the following. 

\subsubsection{Slit spectroscopy with {\it AKARI} IRC/NIR}\label{red_nir}
During a single pointed observation with AOT04 grism mode, eight to nine exposure frames of NIR grism spectroscopic (NG) data and one exposure frame of 3.2\,$\mu$m imaging (N3) data are taken with the IRC/NIR \cite[]{ohya07}. 
The dark current is measured in one frame each at the first and last parts of the pointed observation. 
A single exposure frame consists of one short-exposure and one long-exposure images. 
In the present study, only the long exposure data are used. 
The dark image for each NIR observation is obtained by averaging three long-exposure images of the dark current, which are collected from the adjacent pointed observations including itself to correct for any high-energy ionizing particle (hereafter cosmic-ray) effects by a 1.5-$\sigma$-clipping method. 
In the present analysis, only the dark current data measured in the first part of each pointed observation are used to avoid the latent image effects and subtracted from the observation images. 

We recognize small shifts in position due to the pointing instability by at most $\sim5\arcsec$ during each pointed observation except for Position 5, where an extraordinary large shift of $\sim15\arcsec$ is found. 
The shift in the direction parallel to the slit is corrected so that the spectra of the same area of the sky are extracted. 
Because the shift in the orthogonal direction is uncorrectable, the exposure frames shifted in the direction perpendicular to the slit by more than a pixel ($\sim1.5\arcsec$) as well as those affected by severe artifacts are discarded, except for Position 5 (see \S 2.3.3). 
The remaining images are averaged taking account of the shift in positions in the direction parallel to the slit by a 1.5-$\sigma$-clipping method to remove cosmic-ray events and the artifacts.

The most critical part in the data-reduction of NG spectra is the removal of artificial patterns as well as the foreground components originating from the zodiacal light and diffuse Galactic emission. 
In NIR observations, pixels saturated by cosmic-ray hits or extremely luminous objects often produce artificial line-like patterns, sometimes termed as "column pull-down" and "multiplexer bleed" \cite[]{piph04}. 
Particularly an artificial line-like pattern in the direction parallel to the slit mimics emission or absorption features in the slit spectrum. 
The position of the artificial line-like structure sometimes differs among different  exposures even in the same pointed observation. 
When an artificial line structure emerges in a certain exposure image, 
it is removed by replacing the data of the affected pixels with those of the unaffected pixels of other exposure images at the same position. 
To estimate the foreground components from the zodiacal and diffuse Galactic emission, 
we employ the datasets obtained at positions off the LMC with AOT04 a;Ns (observation IDs:1500719 and 1500720). 
In these observations the NIR spectra are taken with the prism mode (NP) instead of the grism mode (NG). 
Because the emission at the off-LMC positions are too faint for the observations with the NG mode, we use the data with the NP mode to obtain a reliable foreground spectrum. 
The slit positions of both observations are centered at ($\alpha_{\rm 2000},\delta_{\rm 2000}$) = ($\rm 06^{h}\rm 00^{m}\rm 00.^{s}0,-66^\circ36\arcmin30\arcsec$), which is almost at the same ecliptic latitude ($\beta\sim-90^\circ$) as that of the LMC ($\sim-85^\circ$), but is at $\sim9.7^\circ$ away from the center of the LMC. 
The observation log of the off-LMC position is also given in Table \ref{tbl_obs}.

\subsubsection{Slit spectroscopy with {\it AKARI} IRC/MIR-S}
The data reduction procedures including the dark-current subtraction and the cosmic-ray correction for MIR-S spectroscopic observations are basically the same as those for NIR spectroscopic observations. 
During a single pointed observation with AOT04, four exposure frames of SG1 data, four to five exposure frames of SG2 data, and one exposure frame of 9.0\,$\mu$m imaging (S9W) data are taken with the MIR-S channel. 
The dark current is measured in one frame each in the first and the last parts of the pointed observation. 
A single exposure frame consists of one short-exposure image and three long-exposure images. 
The dark image for each MIR-S observation is obtained by averaging three long-exposure images of the dark current by a 1.5-$\sigma$-clipping method to correct for the cosmic-ray effects. 
In this process, only the dark current data measured in the first part of each pointed observation sequence are used to avoid the latent image effects. 
The same shifts in position as those recognized in the NIR data are expected in the MIR-S observations because the same field-of-view is shared by the NIR and MIR-S channels and they are corrected in the same manner as in the NIR data. 
The shift in position among three long-exposure images taken in a single frame is negligible in most cases and, therefore, they are averaged by a 1.5-$\sigma$-clipping method to remove the cosmic-ray effects.

The subtraction of the foreground components (zodiacal light and diffuse Galactic emission) is a more serious problem in the data reduction of MIR spectroscopy than in NIR. 
In addition, the MIR detector suffers scattered light originating in the scattering within the detector array \citep[e.g.,][]{sako07}. 
To estimate the foreground emission and the scattered light component, the SG1 and SG2 spectra collected at the position off the LMC (Observation IDs: 1500719 and 1500720) are used. 
The spectrum at off-position is obtained by averaging the two observations and is subtracted from the spectra of the target positions. 
The MIR-S spectra are basically dominated by the zodiacal light and thus the scattered light component of the off-position spectrum is almost the same as that in the spectra of the target. 
Therefore, the subtraction of the off-position spectrum works effectively not just to remove the foreground emission but also to correct for possible artifacts and greatly improves the resultant spectra. 

\subsubsection{Continuous spectra from NIR to MIR}
Using the spectral response function of each module, we obtain NG, SG1 and SG2 segmental spectra at the same area of the sky except for positions 5 and 8-1 (see below). 
Each segmental spectrum is truncated at the wavelengths where the S/N becomes low: NG is truncated at 2.55\,$\mu$m and 4.9\,$\mu$m, SG1 at 5.5\,$\mu$m and 7.9\,$\mu$m and  SG2 at 7.9\,$\mu$m and 13.4\,$\mu$m. 
Then, the correction for the slit efficiency for extended sources \cite[]{sako08} is applied and continuous spectra from 2.55 to 13.4\,$\mu$m are obtained with a small gap between 4.9 and 5.5\,$\mu$m. 
Because of the severe artifacts (column-pulldown) the NG spectrum is truncated at 3.8\,$\mu$m and 4.5\,$\mu$m for Positions 2 and 5, respectively. 
Note that although there is a small gap between the NG and SG1 segments, all the NG, SG1, and SG2 segmental spectra are smoothly connected to each other without scaling, suggesting that the subtraction procedure of the foreground emission and scattered light works well and reliable spectra are obtained. 

For Positions 5 and 8-1, we cannot obtain segmental spectra precisely at the same region of the sky between SG1 and SG2 because the SG1 and SG2 observations are not carried out simultaneously during a pointed observation. 
A relatively large positional shift ($\sim15\arcsec$) recognized during the pointed observation for Position 5 prevents us from obtaining the SG1 and SG2 spectra from the same region of the sky. 
For Position 8-1, the positional stability during the pointed observation is almost the same as the other observations. 
However, a small shift in the position of the point-like source in the slit of 5$\arcsec$ width between the SG1 and SG2 observations changes the source flux to some extent and makes a small difference in the flux level between the SG1 and SG2 spectra. 
Since the positional stability was better when the SG2 spectrum was taken than the SG1 spectrum for the observations of both positions, only the NG data taken simultaneously with the SG2 are used for the data at both positions. 
Then, the SG1 spectrum is scaled to match with the SG2 spectra in the spectral region of 7.3 to 7.9\,$\mu$m. 
The scaling factors are 0.9 and 0.8 for Position 5 and Position 8-1, respectively. 
The scaling of the SG1 spectrum is taken into account in the derivation of the band intensity ratios (\S 3) and does not make serious effects on the following discussion.

\section{Results}

The resultant spectra toward our target positions are shown in Figure \ref{fig_lmc_raw}. 
The UIR bands are clearly seen at 3.3, 6.2, 7.7, 8.6 and 11.3\,$\mu$m in every spectrum. 
A weak feature around 5.70\,$\mu$m is also seen at Positions 7 and 8-A \cite[]{boer09}.  
The hydrogen recombination line of Br$\alpha$ 4.05\,$\mu$m is detected in four of the spectra, and the hydrogen recombination lines of Br$\beta$ 2.63\,$\mu$m, Pf$\gamma$ 3.75\,$\mu$m, and Pf$\beta$ 4.65\,$\mu$m as well as the forbidden lines of [\ion{Ar}{2}] 6.98\,$\mu$m, [\ion{Ar}{3}] 8.99\,$\mu$m, [\ion{S}{4}] 10.51\,$\mu$m and [\ion{Ne}{2}] 12.81\,$\mu$m are detected in three of the spectra. 
We note that a small bump seen around at 9.6\,$\mu$m is an artifact, originating from the latent of the S9W exposure frame taken just before the SG2 exposure frames. 
Therefore, the spectral data from 9.4 to 9.8\,$\mu$m are not used in the following model fit, and we cut 9.4--9.8\,$\mu$m from spectra in Figure \ref{fig_lmc_raw}.

To derive the intensity of each UIR band and emission line, we fit the observed spectra with  
\begin{equation}
F_\lambda(\lambda)=\sum_{k=0}^{5}a_k \cdot \lambda^{k} + \sum_{k_l=1}^{14}b_{k_l} \cdot \frac{( \gamma_{k_l}/2)^2}{ (\lambda-\lambda_ {k_l})^2+( \gamma_{k_l}/2 )^2 }+\sum_{k_g=1}^{11}c_{k_g} \cdot \exp \displaystyle \biggl[ - \frac{1}{2} \cdot \frac{(\lambda-\lambda_{k_g})^2}{\gamma_{k_g}^2/(8\cdot \rm ln2)} \biggr], 
\label{fit}
\end{equation}
where $\lambda$ is the wavelength. 
The first term represents the continuum, which is modeled with a polynomial function of the 5$^{\rm th}$ order and is constrained to be non-negative. 
The second and third terms correspond to the UIR bands and the emission lines, respectively. 
For the UIR bands, we include 17 components centered 
at 3.30, 3.41, 3.46, 3.51, 3.56, 5.70, 6.22, 6.69, 7.60, 7.85, 8.33, 8.61, 10.68, 11.23, 11.33, 11.99, 12.62\,$\mu$m according to \cite{smit07}.  
Except for the 3.46, 3.51 and 3.56\,$\mu$m components 
the UIR band components have band widths larger than or similar to the spectral resolution of {\it AKARI}/IRC and are thus modeled with Lorentzian profiles (the second term), 
where $\lambda_{k_l}$ is the center wavelength, $\gamma_{k_l}$ is the FWHM and $b_{k_l}$ is the height of each component. 
As for the 3.30, 3.41, 5.70, 6.22, 6.69, 7.60, 7.85, 8.33, 8.61, 10.68, 11.23, 11.33, 11.99, 12.62 components, $\gamma_{k_l}$ is fixed to the best-fit value obtained for the spectrum that has the highest S/N ratio with the spectral resolution of {\it AKARI}/IRC as the minimum value. 
The adopted value of $\gamma_{k_l}$ for each component is summarized in Table \ref{UIR_l}. 
Only the height $b_{kl}$ is left as a free parameter.  
The integrated intensity of each component is calculated as $\pi b_{k_l} \cdot \gamma_{k_l}/2$. 
In the following analysis, the 7.7\,$\mu$m band is defined as a combination of the 7.60 and 7.85\,$\mu$m components,
and the 11.3\,$\mu$m band as a combination of the 11.23 and 11.33\,$\mu$m components. 
We note that the 12.6\,$\mu$m band is defined as one component of the 12.62\,$\mu$m component, because the red-wing component, such as the 12.69\,$\mu$m component suggested by \cite{smit07}, is not detected at a significant level in all the spectra due to the poor S/N and the low spectral resolution. 

The 3.46, 3.51 and 3.56\,$\mu$m components as well as the emission lines, which have the band widths smaller than the spectral resolution of {\it AKARI}/IRC, are modeled with Gaussian profiles (the third term), 
where $\lambda_{k_g}$ is the center wavelength, $\gamma_{k_g}$ is the FWHM and $c_{kg}$ is the height of each component. 
In the fit, $\gamma_{k_g}$ is fixed to match with the spectral resolution of {\it AKARI}/IRC at the corresponding segment. 
Only the height $c_{kg}$ is a free parameter in the fitting. 
The integrated intensity of each Gaussian component is given by $(\pi/\rm ln2)^{1/2} \cdot c_{k_g} \cdot \gamma_{k_g} /2$. 
The adopted values of $\gamma_{k_g}$ for the 3.46, 3.51 and 3.56 $\mu$m components and the emission lines are summarized in Tables \ref{UIR_g} and \ref{line_g}, respectively. 

The best-fit model spectra given by Eq. (\ref{fit}) are plotted together with the observed spectra in Figure \ref{fig_lmc_spec_fit}.  The residual spectra are also plotted in the lower panel of each plot. 
The derived intensities of the major UIR bands at 3.3, 6.2, 7.7, 8.6, 11.3, and 12.6\,$\mu$m and the emission lines of Br$\alpha$ , Br$\beta$ , Pf$\gamma$ , Pf$\beta$,  [\ion{Ar}{2}]6.98, [\ion{Ar}{3}]8.99, [\ion{S}{4}]10.51, and [\ion{Ne}{2}]12.81 are summarized in Tables \ref{UIR_str} and \ref{el_str}, respectively, where those detected with more than 2$\sigma$ are indicated. 
The uncertainties in the intensity are estimated from the fitting errors taking account of the observational uncertainties.

%\subsection{Variation in spectra}
\label{sec:var}
As shown in Table \ref{el_str}, the 3.3, 6.2, 7.7, 8.6, 11.3, and 12.6\,$\mu$m bands are detected at every target position except for the 12.6\,$\mu$m band at Position 5, which shows the faintest emission. 
The 5.70\,$\mu$m band are detected only at Positions 7 and 8-A. 
On the other hand, as shown in Table \ref{el_str}, all of the hydrogen recombination lines of Br$\alpha$ 4.05\,$\mu$m, Br$\beta$ 2.63\,$\mu$m, Pf$\gamma$ 3.75\,$\mu$m, and Pf$\beta$ 4.65\,$\mu$m and the fine structure lines from the ionized gas of [\ion{Ar}{2}] 6.98\,$\mu$m, [\ion{Ar}{3}] 8.99\,$\mu$m, [\ion{S}{4}] 10.51\,$\mu$m and [\ion{Ne}{2}] 12.81\,$\mu$m are detected at Positions 7, 8-1 and 8-2, 
whereas none of them are detected at Positions 1, 2, 3, 4, 5 and 6 except for Br$\alpha$ detected barely at Position 5. 
Taking account of the high ionization potentials of  27.63 eV, 34.83 eV and 21.56 eV to form Ar$^{2+}$, S$^{3+}$ and Ne$^{+}$ \cite[]{alle73}, respectively, Positions 7, 8-1 and 8-2 are exposed to the hard incident radiation field powered by young massive stars and are associated with \ion{H}{2} regions.  
This view is consistent with the radiation field conditions suggested by the {\it IRAS} and {\it AKARI} colors.  
Based on the characteristics of the observed spectra, we classify the targets into two groups: 
"Group A", which includes Positions 1, 2, 3, 4, 5 and 6 and 
"Group B", to which Positions 7, 8-1 and 8-2 belong. 
The members of Group A are supposed to be exposed to incident radiation fields of weak to moderate intensities and consist mostly of molecular clouds and PDRs. 
Group B members are all associated with \ion{H}{2} regions.

%\subsection{Extinction effects on the UIR band ratios}
%\label{sec_extinction}
We investigate the effects of extinction on the spectra observed at the present target positions  according to \cite{doba08}. 
The visual extinction $A_V$ toward the present target positions ranges from 0.0 to 2.5 as shown in the last row of Table \ref{tbl_obs}. 
We assume the "LMC avg" extinction curve provided by \cite{wedr01} to estimate the infrared extinction. 
We also estimate the value of $A_V$ from the observed ratio of Br$\beta$ to Br$\alpha$ at Positions 7, 8-1, and 8-2, assuming the Case B condition of $T_{\rm e}=10^4{\rm K}$ and $n_{\rm e}=10^4 {\rm cm^{-3}}$ \citep{sthu95}. 
The UIR band intensities corrected with $A_V$ provided by \cite{doba08} differ from those corrected with $A_V$ estimated from the Case B condition by less than 10\,\% at Positions 7, 8-1, and 8-2. 
 We adopt those corrected with $A_V$ provided by \cite{doba08} at these positions for consistency with the other target positions. 
The effect of extinction correction on the UIR band ratios is small ($<15\,\%$), and does not affect the following discussion. 

Next, we evaluate a contribution from the unresolved emission line Pf$\delta$ at 3.30\,$\mu$m to the 3.3\,$\mu$m band, and that from Pf$\alpha$ at 7.46\,$\mu$m to the 7.7\,$\mu$m band. 
We assume that the intensities of  Pf$\delta$ and Pf$\alpha$ are equal to 9.3\% and 30.1\% of that of Br$\beta$, respectively, according to the Case B condition and subtract them from the intensity of the 3.3\,$\mu$m and 7.7\,$\mu$m bands. 
In the present target positions, the contribution from Pf$\delta$ to the 3.3\,$\mu$m band is less than 10\% and that from Pf$\alpha$ to the 7.7\,$\mu$m band is less than 3\%, both of which are similar to the measurement uncertainties and thus do not affect the results. 
The intensity for which these corrections are applied  is also listed in the lower row for each position in Tables \ref{UIR_str} and \ref{el_str}. 
The values of the corrected UIR band ratios are listed in Table \ref{UIR_ratio_e_c}. 
The effect of these corrections on the values of the UIR band ratios is less than $\sim10\%$ for all the targets. 
We estimate the uncertainties in $I_{\rm 6.2\,\mu m}$ and $I_{\rm 7.7\,\mu m}$ as the difference between the intensities with and without scaling, which dominates over the fitting errors. 

Figures \ref{UIRratio_IRAS}a--f show the plots of the corrected UIR band ratios of the $I_{\rm 3.3\,\mu m}$/$I_{\rm 11.3\,\mu m}$, $I_{\rm 6.2\,\mu m}$/$I_{\rm 11.3\,\mu m}$, $I_{\rm 7.7\,\mu m}$/$I_{\rm 11.3\,\mu m}$, $I_{\rm 8.6\,\mu m}$/$I_{\rm 11.3\,\mu m}$, $I_{\rm 12.6\,\mu m}$/$I_{\rm 11.3\,\mu m}$, and $I_{\rm 6.2\,\mu m}$/$I_{\rm 7.7\,\mu m}$ against the {\it IRAS} color of $I_{\rm 25\,\mu m}$/$I_{\rm 12\,\mu m}$. 
Figures \ref{UIRratio_IRAS}g--l plot those ratios against the {\it AKARI} color of $I_{\rm L24}$/$I_{\rm S11}$. 
Group A forms a sequence with a positive slope in the plots of $I_{\rm 3.3\,\mu m}$/$I_{\rm 11.3\,\mu m}$, $I_{\rm 6.2\,\mu m}$/$I_{\rm 11.3\,\mu m}$, $I_{\rm 7.7\,\mu m}$/$I_{\rm 11.3\,\mu m}$ and $I_{\rm 8.6\,\mu m}$/$I_{\rm 11.3\,\mu m}$ against the {\it IRAS} and {\it AKARI} colors, while Group B does not follow the sequence in the plots of these band ratios. 
The $I_{\rm 12.6\,\mu m}$/$I_{\rm 11.3\,\mu m}$ and $I_{\rm 6.2\,\mu m}$/$I_{\rm 7.7\,\mu m}$ ratio does not show a systematic trend with the colors either for Groups A or B.  
Little variation found in the 6.2\,$\mu$m to the 7.7\,$\mu$m band ratio is similar to the trend seen in external galaxies \citep{gall08}. 
The present observation shows that the 3.3\,$\mu$m band is weak in Group B compared to Group A. 
The results are discussed in the next section. 

\section{Discussion}
\subsection{Variation in the UIR band ratios}
\label{sec:inte} 
Studies based on quantum chemical calculations as well as laboratory experiments show that 
the ionization fraction and the molecular size of PAHs 
are the primary factors to affect the relative intensity of the UIR band features \citep[e.g.,][]{alla99,drli07,tiel08}. 
The ionization fraction of PAHs is determined by  the balance between the photo-ionization and the recombination with ambient electrons and thus it is expected to be controlled by $G_{\rm 0} \cdot T_{\rm g}^{1/2}/n_{\rm e}$ , where $G_{\rm 0}$ is the intensity of the radiation field in units of the solar vicinity value  \citep[$1.6 \times 10^{-6}$\,W\,m$^{-2}$,][]{habi68}, $n_{\rm e}$ is the electron density, and $T_{\rm g}$ is the gas temperature \citep{bati94,bake01}. 
Higher $G_{\rm 0}$/$n_{\rm e}$ ratios favor larger fractions of positively ionized PAHs \cite[]{bake01}. 
Laboratory experiments and theoretical studies show that 
the ionization of PAHs enhances the intensity of the band features in the 6--9\,$\mu$m region relative to the features in the 11--14\,$\mu$m region \citep{defr93,szva93,alla99,bake01}. 
Therefore, the ionized-to-neutral band ratios (e.g. , $I_{\rm 6.2\,\mu m}$/$I_{\rm 11.3\,\mu m}$,  $I_{\rm 7.7\,\mu m}$/$I_{\rm 11.3\,\mu m}$,  $I_{\rm 8.6\,\mu m}$/$I_{\rm 11.3\,\mu m}$) are supposed to indicate the ionization conditions of PAHs. 
Observational studies report 
variations in $I_{\rm 7.7\,\mu m}$/$I_{\rm 11.3\,\mu m}$ and $I_{\rm 8.6\,\mu m}$/$I_{\rm 11.3\,\mu m}$ 
within a reflection nebula along the distance from the central star \citep{brte05,jobl96}, 
among Herbig Ae/Be stars with different spectral types \cite[]{sloa05}, 
between the interarm and arm regions of the star-forming galaxy NGC6949 \cite[]{sako07}, and 
between the inner and the outer Galaxy \cite[]{sako04}.   
They are reasonably interpreted by the difference in the ionization conditions of the band carriers. 

While very small PAHs ($n_{\rm C} <\sim 10^2$, where $n_{\rm C}$ is the number of carbon atoms in the PAH) radiate strongly at 3.3\,$\mu$m, large PAHs radiate mostly at longer wavelengths: PAHs as large as $n_{\rm C}$ $\sim$ 10$^2$--10$^3$ efficiently convert the absorbed energy to the 6.2, 7.7 and 8.6\,$\mu$m bands and PAHs as large as $n_{\rm C}$ $\sim$ 4000 to the 11.3\,$\mu$m band \citep{schu93,drli07}. 
The 6.2\,$\mu$m and 7.7\,$\mu$m bands are attributed to stretching modes of C-C bonds, while the 3.3\,$\mu$m, the 8.6\,$\mu$m and the 11.3\,$\mu$m bands are attributed to stretching modes, in-plane bending modes, and out-of-plane bending modes of C-H bonds, respectively \citep{alla89}. 
Therefore the ratios of the short-to-long wavelength UIR bands from the same bonds (e.g., $I_{\rm 3.3\,\mu m}$/$I_{\rm 11.3\,\mu m}$,  $I_{\rm 6.2\,\mu m}$/$I_{\rm 7.7\,\mu m}$) can be used to infer  the average temperature or the size distribution of PAHs \citep[e.g.,][]{demu90,sale10,boer10}. 
The NIR to MIR spectra we discuss here are obtained from the same region in the sky and thus we can discuss the band intensity ratios of the major UIR bands at 3.3, 6.2, 7.7, 8.6 and 11.3\,$\mu$m concurrently.

As described in \S \ref{sec:var}, 
Group A forms a sequence on the diagrams of the {\it IRAS} and {\it AKARI} colors v.s. the ionized-to-neutral UIR band ratios: $I_{\rm 6.2\,\mu m}$/$I_{\rm 11.3\,\mu m}$, $I_{\rm 7.7\,\mu m}$/$I_{\rm 11.3\,\mu m}$ and $I_{\rm 8.6\,\mu m}$/$I_{\rm 11.3\,\mu m}$. 
The sequence suggests that a larger fraction of PAHs is ionized as the radiation field becomes stronger. 
However, Group B, whose radiation fields are much stronger and harder than those of Group A, does not follow the sequence. 
This can be attributed to the lower ionization fraction of PAHs due to an increase in the recombination under the high electron density in \ion{H}{2} regions relative to molecular clouds or PDRs. \cite{papo05} argues that hydrogen impact might also play a role in the excitation of PAHs in PDRs and molecular clouds.

There is a positive correlation between {\it IRAS} and {\it AKARI} colors and the short-to-long wavelength UIR band ratio, $I_{\rm 3.3\,\mu m}$/$I_{\rm 11.3\,\mu m}$ in Group A.  
This can be interpreted in terms of an increase in the excitation temperature of PAHs with the {\it IRAS} and {\it AKARI} colors in Group A. 
Destruction of PAHs is expected to be inefficient in environments of Group A \citep{mice10b,mice10a,mice11} and the size distribution of PAHS does not change considerably. 
Thus the excitation temperature of PAHs is mainly controlled by the hardness of the incident radiation field, but not by its intensity. 
On the other hand, the {\it IRAS} and {\it AKARI} colors indicate its intensity, but not the hardness directly \citep{sako06,onak07}. 
The correlation seen in Figure \ref{UIRratio_IRAS} thus indicates that the incident radiation field becomes harder as the intensity becomes larger in Group A.  
This is a reasonable consequence of strong incident radiation fields, for which a contribution from young massive stars becomes dominant. 
The present observations indicate this trend explicitly based on the NIR to MIR UIR band ratios. 
Harder incident radiation fields also increase the $I_{\rm 6.2\,\mu m}$/$I_{\rm 11.3\,\mu m}$, $I_{\rm 7.7\,\mu m}$/$I_{\rm 11.3\,\mu m}$, and $I_{\rm 8.6\,\mu m}$/$I_{\rm 11.3\,\mu m}$. 
We show in \S \ref{sec:comp} that the ionization fraction is the major factor for the increase in these ratios with a minor contribution from the hardness of the incident radiation field. 

Group B does not follow the trend of Group A and shows the weaker 3.3\,$\mu$m band. 
It can be attributed to destruction of very small PAHs ($n_{\rm C} <\sim 10^2$) inside \ion{H}{2} regions, which are supposed to dominantly contribute to the 3.3\,$\mu$m band \citep{schu93,drli07}. 
\cite{mice10b} suggest that electron collisions dominate in the destruction of PAHs in a hot plasma for gas temperatures between $\sim 3 \times 10^4$ and $\sim 10^6$\,K, where smaller PAHs are more efficiently destroyed than larger ones. 
The threshold temperatures depend sensitively on the assumed electronic dissociation energy. 
The difference of $I_{\rm 3.3\,\mu m}$/$I_{\rm 11.3\,\mu m}$ between Group A and B may thus result from the destruction by electron collisions, if it is a dominant process of PAH destruction in Group B. 
The absence of a systematic trend with the IR colors may be partly due to the small number of the data in Group B.

\subsection{Diagnostic diagram of the radiation field condition}
\subsubsection{The diagnostic diagram of the UIR band ratios}
\label{sec:diagram}
According to the discussion in the previous subsections and the results presented in \S \ref{sec:var}, 
we explore possible diagnostic diagrams to investigate the physical conditions of the ISM by means of the UIR band ratios. 
The ratio $I_{\rm 3.3\,\mu m}$/$I_{\rm 11.3\,\mu m}$ is found to be a good indicator of the size distribution of PAHs, whereas $I_{\rm 6.2\,\mu m}$/$I_{\rm 11.3\,\mu m}$, $I_{\rm 7.7\,\mu m}$/$I_{\rm 11.3\,\mu m}$, or $I_{\rm 8.6\,\mu m}$/$I_{\rm 11.3\,\mu m}$ has been indicated to be a measure of the ionized fraction of PAHs as discussed above. 
The 8.6\,$\mu$m band is typically weak and situated on the shoulder of the strong 7.7\,$\mu$m band due to the limited resolution provided by the IRC. 
Thus $I_{\rm 8.6\,\mu m}$/$I_{\rm 11.3\,\mu m}$ is less reliable than the other two ratios and will not be considered. 
Taking account of these, we investigate two-band-ratio plots of $I_{\rm 3.3\,\mu m}$/$I_{\rm 11.3\,\mu m}$ v.s. $I_{\rm 7.7\,\mu m}$/$I_{\rm 11.3\,\mu m}$ or $I_{\rm 6.2\,\mu m}$/$I_{\rm 11.3\,\mu m}$ as shown in Figures \ref{UIRratio_UIRratio}a and b. 

In Figure \ref{UIRratio_UIRratio}a where $I_{\rm 3.3\,\mu m}$/$I_{\rm 11.3\,\mu m}$ is plotted against $I_{\rm 7.7\,\mu m}$/$I_{\rm 11.3\,\mu m}$, Group A forms a sequence from the bottom-left to the top-right. 
This sequence can be interpreted in terms of the change in the spectrum of the  incident radiation field together with the change in the ionization fraction of PAHs.
Group B, however, does not follow the sequence and is located distinctly from Group A. 
Figure \ref{UIRratio_UIRratio}b shows a plot of $I_{\rm 3.3\,\mu m}$/$I_{\rm 11.3\,\mu m}$ v.s. $I_{\rm 6.2\,\mu m}$/$I_{\rm 11.3\,\mu m}$. 
Group A forms a sequence similar to the plot of $I_{\rm 3.3\,\mu m}$/$I_{\rm 11.3\,\mu m}$ v.s. $I_{\rm 7.7\,\mu m}$/$I_{\rm 11.3\,\mu m}$ and Group B data are also separated from Group A. 
The separation of Group B from Group A is more distinct in Figure \ref{UIRratio_UIRratio}a than in Figure \ref{UIRratio_UIRratio}b. 
Lower left points of Group B in Figure \ref{UIRratio_UIRratio}b may overlap with the sequence of Group A if the sequence continues down to the lower ratios and would become difficult to be distinguished from Group A.
The difference between the two plots comes from a slight decrease in the 6.2\,$\mu$m band intensity relative to the 7.7\,$\mu$m band in Group B, in particular in Position 8-1 and 8-2 spectra. 
While the $I_{\rm 6.2\,\mu m}$/$I_{\rm 7.7\,\mu m}$ ratio of Group A does not show a clear dependency on the intensity and/or hardness of the radiation field (Figures \ref{UIRratio_IRAS}e and j), the average value of the $I_{\rm 6.2\,\mu m}$/$I_{\rm 7.7\,\mu m}$ ratio of Group B (0.303$\pm$0.053) is slightly smaller than that of Group A (0.375$\pm$0.051). 
This may suggest that even medium-sized PAHs are destroyed in \ion{H}{2} regions in Group B \citep{schu93, drli01, sale10}. 
Alternatively, de-hydrogenation might also explain this trend. 
Since C-H in-plane bending modes contribute to part of the 7.7\,$\mu$m band, H-loss reduces the intensity of the 7.7\,$\mu$m band. 
However, there is a large scatter in $I_{\rm 6.2\,\mu m}$/$I_{\rm 7.7\,\mu m}$ among each group. 
We cannot rule out a possibility that there are other causes for the variation in $I_{\rm 6.2\,\mu m}$/$I_{\rm 7.7\,\mu m}$.

In the present results, Groups A and B can be distinguished more clearly in the diagram of $I_{\rm 3.3\,\mu m}$/$I_{\rm 11.3\,\mu m}$ v.s. $I_{\rm 7.7\,\mu m}$/$I_{\rm 11.3\,\mu m}$ than that of  $I_{\rm 3.3\,\mu m}$/$I_{\rm 11.3\,\mu m}$ v.s. $I_{\rm 6.2\,\mu m}$/$I_{\rm 11.3\,\mu m}$. 
It suggests a potential of the diagram of $I_{\rm 3.3\,\mu m}$/$I_{\rm 11.3\,\mu m}$ v.s. $I_{\rm 7.7\,\mu m}$/$I_{\rm 11.3\,\mu m}$ as a diagnostic tool for the radiation field conditions. 
The 6.2\,$\mu$m band is usually thought as a more reliable indicator than the 7.7\,$\mu$m band, since the 6.2\,$\mu$m consists of a single component, while the 7.7\,$\mu$m contains more than one component \citep{peet02,smit07}. 
The present results indicate that the 7.7\,$\mu$m band might be a better indicator than the 6.2\,$\mu$m band as far as the ionization fraction of PAHs are concerned probably because it is less sensitive to the variation in the size distribution. 
The number of Group B targets is small, however, and thus we need further investigations to ensure if this conclusion can be applied to the general ISM with a wide range of physical conditions.  

\cite{verm02} investigate the variations in the MIR UIR bands among the \ion{H}{2} regions of the LMC based on ISOPHOT/PHT-S observations and find that the $I_{\rm 6.2\,\mu m}$/$I_{\rm 11.3\,\mu m}$ and $I_{\rm 7.7\,\mu m}$/$I_{\rm 11.3\,\mu m}$ ratios are systematically smaller in HII regions in 30 Doradus and the Small Magellanic Cloud (SMC) than in Galactic and non-30 Doradus regions. 
They suggest that the segregation may be attributed to the difference in the molecular structure of PAHs, proposing that compact PAHs may dominate in the SMC and 30 Doradus regions. 
The present targets do not include 30 Doradus itself. 
Also none of them have the incident radiation fields of the intensity similar to the 30 Doradus region. 
The 11.3\,$\mu$m band is due to solo C-H out-of-plane bending modes and probes long straight edges, whereas the 12.6\,$\mu$m band is due to trio C-H out-of-plane modes and probes corners. 
The contribution from compact PAHs enhances the 11.3\,$\mu$m band relative to the 12.6\,$\mu$m band. 
The different compactness of PAHs is supposed to appear more distinctly in the $I_{\rm 12.6\,\mu m}$/$I_{\rm 11.3\,\mu m}$ ratio than in the $I_{\rm 3.3\,\mu m}$/$I_{\rm 11.3\,\mu m}$. 
The ratio of $I_{\rm 12.6\,\mu m}$/$I_{\rm 11.3\,\mu m}$ is $0.40\pm 0.10$ and $0.44\pm0.03$ for Group A and B, respectively.  
Although the intensity of the 12.6\,$\mu$m band has a large uncertainty, the ratio does not show a distinct difference as seen in $I_{\rm 3.3\,\mu m}$/$I_{\rm 11.3\,\mu m}$, suggesting that the molecular structure does not change appreciably among the present targets and is not the major factor for the difference in $I_{\rm 3.3\,\mu m}$/$I_{\rm 11.3\,\mu m}$. 

Recently \cite{hony11} also investigate the variations in the MIR UIR bands based on {\it Spitzer}/IRS observations of the LMC. 
Their targets include molecular clouds, PDRs and \ion{H}{2} regions as in the present study, but the number of the targets is much larger. 
They find that the ratio $I_{\rm 7.7\,\mu m}$/$I_{\rm 11.3\,\mu m}$ is mostly controlled by the ionization fraction of PAHs and weakly depends on the [\ion{Ne}{2}]/[\ion{Ne}{3}] line ratio, which they interpreted as the indicator of hardness of the incident radiation field. 
The present results show a similar trend, but also suggest a systematic trend with the hardness, once molecular clouds and PDRs are separated from \ion{H}{2} regions by use of the ratio $I_{\rm 3.3\,\mu m}$/$I_{\rm 11.3\,\mu m}$.  
The sequence seen for Group A suggests that the excitation of PAHs as indicated by $I_{\rm 3.3\,\mu m}$/$I_{\rm 11.3\,\mu m}$ is enhanced with the ionization fraction, which is indicated by $I_{\rm 7.7\,\mu m}$/$I_{\rm 11.3\,\mu m}$ or $I_{\rm 6.2\,\mu m}$/$I_{\rm 11.3\,\mu m}$. 
Thus the ratio $I_{\rm 3.3\,\mu m}$/$I_{\rm 11.3\,\mu m}$ is a useful measure for the incident radiation field conditions for Group A targets. 
This interpretation suggests that the data at the lower left in Figure \ref{UIRratio_UIRratio} are objects in an early stage of the cloud evolution
and those at the upper right are more evolved PDR-type objects. 
Further observations are important to confirm the interpretation. 
In both plots of Figure \ref{UIRratio_UIRratio}, the ratio $I_{\rm 3.3\,\mu m}$/$I_{\rm 11.3\,\mu m}$ plays a crucial role not only to separate Group A from B, but also to estimate the  incident radiation field conditions of the target in Group A.

\subsubsection{Comparison with model}
\label{sec:comp}
The previous subsection suggests possible diagnostic diagrams based on the UIR band ratios and qualitative interpretation is given. 
In this subsection we employ simple models of PAH emission and investigate the observed diagrams quantitatively.

To derive the intensity of the UIR bands emitted from a mixture of neutral and ionized PAHs of various sizes, 
we employ a simple theoretical model of infrared PAH emission following \cite{schu93}. 
The A-coefficients of the UIR bands in 3--20\,$\mu$m are calculated from the infrared cross sections of neutral and ionized PAHs recently provided by \cite{drli07} and the internal energy to temperature relation provided by \cite{drli01} is adopted \citep[see also][]{boer10,baus10}. 
The number of carbon atoms in PAHs is assumed to be distributed between $n_{\rm C}^{\rm min}$ and $n_{\rm C}^{\rm max}$, and the size distribution of PAHs is assumed to be given by the same power-law function as graphite grains. 
Details on the model calculation are given in Appendix \ref{ap:cal}. 

The UIR band ratios of $I_{\rm 3.3\,\mu m}$/$I_{\rm 11.3\,\mu m}$ and $I_{\rm 7.7\,\mu m}$/$I_{\rm 11.3\,\mu m}$ are calculated for different temperatures of the heating source $T_*$ and ionization fractions of PAHs $f_{\rm i}$. 
Two sets of ($n_{\rm C}^{\rm min}$, $n_{\rm C}^{\rm max}$) are calculated in the following analysis: Case I assumes ($n_{\rm C}^{\rm min}$, $n_{\rm C}^{\rm max}$) = (20, 4000). 
The UIR band ratios of $I_{\rm 3.3\,\mu m}$/$I_{\rm 11.3\,\mu m}$ and $I_{\rm 7.7\,\mu m}$/$I_{\rm 11.3\,\mu m}$ calculated for various $T_*$ and $f_{\rm i}$ are plotted in Figure \ref{model_ratio}a. 
Compared with the locations of the Group A data points on the plot, the bottom-left edge of the sequence corresponds to $T_*\sim$7500\,K and $f_{\rm i}\sim$30\,\%, while the top-right end of the sequence to that with $T_*\sim$12000\,K and $f_{\rm i}\sim$60\,\%. 
The range of the effective temperature indicated is compatible with the effective temperature of A-type main sequence stars of 7000--10000K. 
Thus the model calculations support the interpretation in \S \ref{sec:inte} and the observed sequence of Group A can be accounted for by the increase of the effective temperature of the heating source, which increases the excitation, accompanied by the increase in the ionization fraction of PAHs without the change in the size distribution.
The observed data of Group B, on the other hand, are distributed in a region corresponding to $T_*\sim$6500--8000\,K and $f_{\rm i}\sim$40--60\,\% for Case I, which is totally inconsistent with the fact that the radiation field of Group B should be much harder than that of Group A. 

Case II assumes ($n_{\rm C}^{\rm min}$, $n_{\rm C}^{\rm max}$)\,=\,(100, 4000). 
The UIR band ratios of $I_{\rm 3.3\,\mu m}$/$I_{\rm 11.3\,\mu m}$ and $I_{\rm 7.7\,\mu m}$/$I_{\rm 11.3\,\mu m}$ calculated for various $T_*$ and $f_{\rm i}$ are plotted in Figure \ref{model_ratio}b. 
The observed data of Group B are distributed in a region corresponding to $T_*\sim$30,000--40,000\,K and $f_{\rm i}\sim$20--40\,\%, which is compatible with the effective temperature of O-type main sequence stars of 30,000--50,000\,K. 
The small $f_{\rm i}$ is consistent with the increase in recombination with electrons in high electron density environments of \ion{H}{2} regions. 
Therefore, the deviation of Group B from the sequence of Group A on the plot of $I_{\rm 3.3\,\mu m}$/$I_{\rm 11.3\,\mu m}$ v.s. $I_{\rm 7.7\,\mu m}$/$I_{\rm 11.3\,\mu m}$ is reasonably accounted for by the difference in the minimum size of PAHs. 

Therefore, the model calculations quantitatively support the interpretation of the trend seen in the plot of $I_{\rm 3.3\,\mu m}$/$I_{\rm 11.3\,\mu m}$ v.s. $I_{\rm 7.7\,\mu m}$/$I_{\rm 11.3\,\mu m}$.
We stress here that an accurate measurement of $I_{\rm 3.3\,\mu m}$/$I_{\rm 11.3\,\mu m}$ ratio should be the key to obtain the average excitation temperature of PAHs, which is determined by the size distribution of PAHs and the hardness of the incident radiation field. 
The present calculation of Case I and II is made only for representative purposes and the choice of $n_{\rm C}^{\rm min}$ is rather arbitrary.  
The value of $I_{\rm 3.3\,\mu m}$/$I_{\rm 11.3\,\mu m}$ is sensitive to the specific heat as well as the size distribution of PAHs that we assume in the calculation. 
The actual values of $T_*$, $f_{\rm i}$, and $n_{\rm C}^{\rm min}$ inferred from the plot depend on the assumed PAH properties such as the size distribution and the specific heat, which are also questionable.

\section{Summary}
We present the results of NIR to MIR slit spectroscopic observations of the diffuse radiation toward nine positions
 with different radiation field conditions in the LMC with {\it AKARI}/IRC. 
We obtain continuous spectra from 2.55 to 13.4\,$\mu$m of the same slit area, 
which allow us to investigate variations in the relative intensity of the UIR bands from 2.55 to 13.4\,$\mu$m emitted from exactly the same region. 

The target positions are selected based on the {\it IRAS} colors of $I_{\rm 25\,\mu m}$/$I_{\rm 12\,\mu m}$ and $I_{\rm 60\,\mu m}$/$I_{\rm 100\,\mu m}$, which indicate star formation activities, to cover a wide range of the intensity of the incident radiation field. 
The {\it AKARI} color of $I_{\rm L24}$/$I_{\rm S11}$ at the present target positions shows a similar trend to that of the {\it IRAS} $I_{\rm 25\,\mu m}$/$I_{\rm 12\,\mu m}$ color, confirming that the selection based on the {\it IRAS} colors is in fact relevant to the purpose of the present study. 

A series of the major UIR bands at 3.3, 6.2, 7.7, 8.6 and 11.3\,$\mu$m are clearly detected in every obtained spectrum, and the hydrogen recombination lines of Br$\alpha$ 4.05\,$\mu$m, Br$\beta$ 2.63\,$\mu$m, Pf$\gamma$ 3.75\,$\mu$m, and Pf$\beta$ 4.65\,$\mu$m and the fine structure lines of [\ion{Ar}{2}] 6.98\,$\mu$m, [\ion{Ar}{3}] 8.99\,$\mu$m, [\ion{S}{4}] 10.51\,$\mu$m and [\ion{Ne}{2}] 12.81\,$\mu$m, signatures of the presence of ionized gas, are detected in some of them. 
According to the ionization gas signatures in the spectra, we classify the present nine target positions into two groups: those without the strong ionized gas signatures (Group A) and those with the signatures (Group B). 
Group A positions are supposed to be in relatively quiescent radiation field environments and those of Group B are in harsh radiation field environments powered by young massive stars. 
This view is consistent with the radiation field conditions suggested by the {\it IRAS} and {\it AKARI} MIR colors. 
 
Group A shows a sequence on the plots of the UIR band ratios of $I_{\rm 3.3\,\mu m}$/$I_{\rm 11.3\,\mu m}$, $I_{\rm 6.2\,\mu m}$/$I_{\rm 11.3\,\mu m}$,  $I_{\rm 7.7\,\mu m}$/$I_{\rm 11.3\,\mu m}$ and $I_{\rm 8.6\,\mu m}$/$I_{\rm 11.3\,\mu m}$ against the {\it IRAS} and {\it AKARI} colors, 
but Group B does not follow the sequence. 
These results can be interpreted in terms of the facts that  
(1) in Group A, PAHs are heated to higher excitation temperatures and their ionization fraction increases as the radiation field becomes harder and stronger  
and that 
(2) in Group B, very small PAHs ($n_{\rm C}<100$) are efficiently destroyed, possibly due to electron collisions, and the ionization of PAHs is suppressed by an increase in the electron density inside \ion{H}{2} regions. 
The present observations also show that the incident radiation field becomes harder as the intensity increases in Group A based on the UIR band ratios and the IR colors. 
There is little variation in $I_{\rm 6.2\,\mu m}$/$I_{\rm 7.7\,\mu m}$ as reported in previous studies and we find no systematic trend against the colors. 

The observed data points of Groups A and B are well separated on the plot of $I_{\rm 3.3\,\mu m}$/$I_{\rm 11.3\,\mu m}$ v.s. $I_{\rm 7.7\,\mu m}$/$I_{\rm 11.3\,\mu m}$ as well as that of $I_{\rm 3.3\,\mu m}$/$I_{\rm 11.3\,\mu m}$ v.s. $I_{\rm 6.2\,\mu m}$/$I_{\rm 11.3\,\mu m}$. 
These trends can be interpreted in the same way as described above, suggesting  a potential of the diagram of $I_{\rm 3.3\,\mu m}$/$I_{\rm 11.3\,\mu m}$ v.s. $I_{\rm 7.7\,\mu m}$/$I_{\rm 11.3\,\mu m}$ as a useful diagnostic tool for the radiation field conditions. 
Simple model calculations support the interpretation quantitatively. 
Further investigation is needed to ensure the applicability to this diagram a wide range of objects. 

The present study shows the importance of the UIR band at 3.3\,$\mu$m. 
The ratio $I_{\rm 3.3\,\mu m}$/$I_{\rm 11.3\,\mu m}$ plays a crucial role not only to separate Group A from B, but also to estimate the incident radiation field conditions of the target in Group A.  
Recently \cite{seok11} report the detection of the 3.3\,$\mu$m UIR band at the supernova remnant (SNR) N49 in the LMC, suggesting the presence of significant processing of PAHs in SNR shocks. 
The 3.3\,$\mu$m band provides significant information on the size distribution and/or the excitation conditions of PAHs.

\acknowledgments

This work is based on observations with {\it AKARI}, a JAXA project with the participation of ESA. 
The authors thank all the members of the {\it AKARI} project and the members of the Interstellar and Nearby Galaxy team for their help and continuous encouragements. 
The ISSA data were obtained from the NASA Astrophysics Data Center (ADC). 
We express our gratitude to K. Dobashi for providing us with the extinction data of the LMC. 
Additionally, we express our gratitude to Y. Ita for providing us with the LMC survey program data by {\it AKARI}. 
We  thank A. Kawamura and Y. Fukui for providing us the LMC 12CO survey data at 2.7 mm taken by the NANTEN millimeter-wave telescope of Nagoya University.
This work is supported in part by a Grant-in-Aid for Scientific Research from the Japan Society of Promotion of Science (JSPS).

\appendix
\section{Appendix}
\label{ap:cal}
The present model calculation basically follows \cite{schu93} with the recent model parameters provided by \cite{drli01} and \cite{drli07} \citep[see also][]{boer10,baus10}. 
The emission intensity due to an IR active fundamental vibrational transition, $i$, from level ($v$-1) to level $v$, in a $j$-type PAH molecule with a total internal vibrational energy $E$ is given by
\begin{eqnarray}
I(j,E,i,v)=h\nu_ivA_{j,i}^{1,0}\frac{\rho_{j,r}(E-vh\nu_i)}{\rho_{j}(E)},
\end{eqnarray}
where $h$ is the Plank constant, $v$ is the vibrational quantum number, $\nu_i$ is the frequency of the emitting mode, $A_{j,i}^{1,0}$ is the Einstein coefficient of the 1 $\to$ 0 transition, $\rho_j(E)$ is the total density of vibrational states at total energy $E$, i.e., the number of ways the energy $E$ can be distributed over all available states, and $\rho_{j,r}(E-vh\nu_i)$ is the density of vibrational states for all modes except the emitting mode at a vibrational energy $E-vh\nu_i$.  
For the Einstein coefficient of the 1 $\to$ 0 transition, $A_{j,i}^{1,0}$ is given by
\begin{equation}
A_{j,i}^{1,0}=\frac{8\pi c}{\lambda_i^4}\sigma_{j,int,i}, 
\end{equation}
where $\lambda_i$ is the wavelength of the emitting mode, $\sigma_{j,int,i}$ is the cross section of the $i$-th mode integrated over the wavelength. 
We adopt the values described in Table 1 of \cite{drli07} as $\sigma_{j,int,i}$ for each type of PAH molecules. 
Here we calculate the model spectra with various parameters to semi-quantitatively compare with the observations and thus adopt a simple thermal approximation. The validity of the thermal approximation has been studied to a large extent and it is shown that the presence of the size distribution alleviates the difference and the effect on the relative band intensities is small enough for the present study 
\citep[]{alla89,schu93,drli01}.

In the thermal approximation, the emitted intensity of a $j$-type molecule with internal energy $E$ in the $i$-th mode, from level $v$ to level ($v$-1), is described by
\begin{equation}
I(j,E,i,v)=h\nu_ivA_{j,i}^{1,0}\exp(-vh\nu_i/kT_j(E))[1-\exp(-h\nu_i/kT_j(E))]^{-1}, \label{taso}
\end{equation}
where $T_j(E)$ is the vibrational excitation temperature of a $j$-type molecule with $E$ and where $k$ is the Boltzmann constant. 
The sum of equation (\ref{taso}) from $v=1$ to $v=\infty$, i.e., the total emitted intensity in the $i$-th mode, is given by
\begin{equation}
I(j,E,i)=h\nu_iA_{j,i}^{1,0}[\exp(h\nu_i/kT_j(E))-1]^{-1}.  \label{taso2}
\end{equation}

In this approximation, the energy-temperature relation for a $j$-type molecule is given by 
\begin{eqnarray}
E(T_j)=\sum_{i=1}^{s}h\nu_i[\exp(h\nu_i/kT_j)-1]^{-1}, 
\end{eqnarray}
where $s$ is the number of vibrational modes of a $j$-type molecule equal to 3 times of atoms minus 6, i.e., $3n_{\rm atom}-6$. 
We derive this relationship, using equations (2)--(8) in \cite{drli01} for each type of PAH molecules. 

The total energy emitted in the $i$-th mode following the absorption of a UV/visual photon of frequency of $\nu$, $f(j,\nu,i)$ is given by
\begin{eqnarray}
f(j,\nu,i)=\int^{h\nu}_0 \frac{I(j,E,i)}{I(j,E)} dE, 
\end{eqnarray}
where $I(j,E,i)/I(j,E)$ is the fraction of the total IR intensity emitted by a $j$-type molecule with the internal energy $E$ in the $i$-th mode. 
Hence, the total emitted intensity in the $i$-th mode of a $j$-type molecule exposed by a star, whose spectrum is approximated with a blackbody, $P(j,T_\ast,i)$ is calculated as
\begin{eqnarray}
P(j,T_\ast,i)=\int^\infty_0 \frac{\sigma_{j,\nu}B_\nu(T_\ast)}{h\nu}f(j,v,i) d\nu, 
\end{eqnarray}
where $\sigma_{j,\nu}$ is the UV/visual absorption cross section of $j$-type molecule, as which we adopt the values described in equations (17)--(20) in \cite{schu93}. 
Then, the flux in the $i$-th mode from a certain size distribution of interstellar PAHs, exposed to a star with $T_\ast$, is given by 
\begin{eqnarray}
F(T_\ast,i)=\sum_{j}n_{\rm PAH}(j)P(j,T_\ast,i), 
\end{eqnarray}
where $n_{\rm PAH}(j)$ is the number density of the $j$-type PAH molecule.

In order to evaluate the effect of depletion of very small PAHs quantitively, we calculate the ratio of $F(T_\ast,i)$ relative to the $F(T_\ast,i$'), the model ratio of the emitted intensity in the $i$-th mode relative to that in the $i'$-th mode in two cases; Case I and Case II (see text). 
In both cases, we assume that  the number density of interstellar PAHs is given by a power-law distribution with the size, i.e., the number of carbon atoms contained by PAH molecules, $n_{\rm C}$. 
If the number of PAH molecules per interstellar H atoms with a radius $a$ between $a+da$ is proportional to $a^{-\alpha}$ and $n_{\rm C} \propto a^\gamma$, the number of PAH molecules per interstellar H atoms with a number of carbon atoms $n_{\rm C}$ between $n_{\rm C}+dn_{\rm C}$ is given by 
\begin{eqnarray}
\frac{dN_{PAH}}{N_H} \propto n_{\rm C}^\frac{1-\alpha-\gamma}{\gamma} dn_{\rm C} . 
\end{eqnarray} 
We adopt $\alpha$=3.5 and $\gamma$=3 as assumed in \cite{schu93}. 
Then the number density is given by 
\begin{eqnarray}
n_{PAH}(j)\propto n_{\rm C}^{-1.833}. 
\end{eqnarray} 
In Case I, the minimum size of PAHs is set to $n_{\rm C}$=20. 
\cite{alla89} suggest that smaller PAHs with $n_{\rm C}$ $<$ 20 are photolytically unstable.
In Case II, we set 
the minimum size of PAHs $n_{\rm C}$=100. 
In both cases, the maximum sizes of PAH molecules is fixed at 4000. 
\cite{drli07} suggest that larger PAHs with $n_{\rm C}$ $>$ 4000 do not contribute to the UIR band features at 3--11\,$\mu$m. 
Then, we run the model twice with the cross sections for neutral PAHs and ionic PAHs provided by \cite{drli07}, respectively, and calculate the ratio of  $F(T_\ast,i)$ from a summation of the spectra of both components by varying the ionization fraction.

\bibliographystyle{apj}
\bibliography{apj-jour,reference}

\begin{deluxetable}{lccccccc}
\tabletypesize{\scriptsize}
%\rotate
\tablecaption{Observation log and parameters\label{tbl_obs}}
\tablewidth{0pt}
\tablehead{
\colhead{Position ID} & \colhead{Date} & \colhead{Obs.ID} & \colhead{AOT} & \colhead{Disperser} &\multicolumn{2}{c}{Center position of the slit} & \colhead{$A_V$\tablenotemark{a}} \\
 &  &  &  &  &  
\colhead{$\alpha_{\rm 2000}$} & \colhead{$\delta_{\rm 2000}$} &
}
\startdata
Position 1 & 2006 October 14 & 1400318.1 & IRC04 b:Ns & NG, SG1 and SG2 & $\rm 05^{h}\rm 38^{m}\rm 20.^{s}$4 & -70$^\circ$07$\arcmin$25$\arcsec$& 1.89  \\
Position 2 & 2006 October 19 & 1400330.1 & IRC04 b:Ns & NG, SG1 and SG2 & $\rm 05^{h}\rm 39^{m}\rm 43.^{s}7$ & -70$^\circ$00$\arcmin$34$\arcsec$& 0.721  \\
Position 3 & 2006 October 2 & 1400346.1 & IRC04 b:Ns & NG, SG1 and SG2 & $\rm 05^{h}\rm 48^{m}\rm 08.^{s}2$ & -69$^\circ$52$\arcmin$52$\arcsec$& 1.90  \\
Position 4 & 2007 May 15        & 1402426.1 & IRC04 b:Ns & NG, SG1 and SG2 & $\rm 05^{h}\rm 25^{m}\rm 55.^{s}4$ & -66$^\circ$10$\arcmin$30$\arcsec$& 0.257  \\
Position 5 & 2006 October 16 & 1400324.1 & IRC04 b:Ns & NG, SG1 and SG2 & $\rm 05^{h}\rm 43^{m}\rm 42.^{s}5$ & -69$^\circ$22$\arcmin$32$\arcsec$& 0.826   \\
Position 6 & 2007 May 18        & 1402422.1 & IRC04 b:Ns & NG, SG1 and SG2 & $\rm 05^{h}\rm 26^{m}\rm 06.^{s}3$ & -68$^\circ$36$\arcmin$04$\arcsec$& 0.630 \\
Position 7 & 2006 October 16 & 1400334.1 & IRC04 b:Ns & NG, SG1 and SG2 & $\rm 05^{h}\rm 39^{m}\rm 17.^{s}67$ & -69$^\circ$30$\arcmin$14$\arcsec$& 0.541 \\
Position 8 & 2006 October 16 & 1400320.1 & IRC04 b:Ns & NG, SG1 and SG2 & $\rm 05^{h}\rm 39^{m}\rm 57.^{s}5$ & -69$^\circ$45$\arcmin$27.5$\arcsec$& 2.11 \\ \hline
%Position 8-1 & 20:56:44 on 2006 October 16 & 1400320.1 & Spectroscopy & IRC04 b:Ns & NG, SG1 and SG2 & $\rm 05^{h}\rm 39^{m}\rm 59.^{s}2$ & -69$^\circ$45'28" & N159-Y4 and N159-O3 \\ 
%Position 8-2 & 20:56:44 on 2006 October 16 & 1400320.1 & Spectroscopy & IRC04 b:Ns & NG, SG1 and SG2 & $\rm 05^{h}\rm 39^{m}\rm 57.^{s}0$ & -69$^\circ$45'20" & N159-Y4 and N159-O3 \\  \hline
LMC-off & 2006 November 10 & 1500719.1 & IRC04 a:Ns & NP, SG1 and SG2 & $\rm 06^{h}\rm 00^{m}\rm 00.^{s}0$ & -66$^\circ$36$\arcmin$30$\arcsec$&- \\ 
LMC-off & 2006 November 10 & 1500720.1 & IRC04 a:Ns & NP, SG1 and SG2 & $\rm 06^{h}\rm 00^{m}\rm 00.^{s}0$ & -66$^\circ$36$\arcmin$30$\arcsec$&-  
\enddata
\tablenotetext{a}{From \cite{doba08}. }
\end{deluxetable}

\begin{deluxetable}{lccc}
\tabletypesize{\scriptsize}
%\rotate
\tablecaption{{\it IRAS} and {\it AKARI} colors\tablenotemark{a} \label{tbl_tgt_imf}}
\tablewidth{0pt}
\tablehead{
\colhead{Position ID} & \colhead{$I_{\rm 25 \mu m}$/$I_{\rm 12\mu m}$} & \colhead{$I_{\rm 60\mu m}$/$I_{\rm 100\mu m}$}  & \colhead{$I_{\rm L24}$/$I_{\rm S11}$}
}
\startdata
Position 1 & 1.01$\pm$0.01 &0.371$\pm$0.004 &  0.58$\pm$0.04 \\
Position 2 & 1.17$\pm$0.01 &0.359$\pm$0.001 &  0.76$\pm$ 0.05 \\
Position 3 & 1.30$\pm$0.05 &0.317$\pm$0.004 &  1.3$\pm$0.1 \\
Position 4 & 1.47$\pm$0.05 &0.435$\pm$0.004 &  1.4$\pm$0.1 \\
Position 5 & 1.67$\pm$0.04 &0.474$\pm$0.006 &  1.5$\pm$0.1 \\
Position 6 & 1.82$\pm$0.02 &0.467$\pm$0.004 &  1.8$\pm$0.1 \\
Position 7 & 4.43$\pm$0.16 &0.751$\pm$0.014 & 6.9$\pm$0.5 \\
Position 8-1 & 4.44$\pm$0.34 &0.730$\pm$0.010 & 7.9$\pm$0.5 \\
Position 8-2 & 4.44$\pm$0.34 &0.730$\pm$0.010 & 6.9$\pm$0.5
\enddata
\tablenotetext{a}{The values are the ratios of the radiances in W m$^{-2}$ Hz$^{-1}$ arcsec$^{-2}$.}
\end{deluxetable}

\begin{deluxetable}{cc}
\tabletypesize{\scriptsize}
%\rotate
\tablecaption{UIR band parameters modeled by Lorentzian profiles \label{UIR_l}}
\tablewidth{0pt}
\tablehead{
 \colhead{$\lambda_{k_l}$ [$\mu$m]} & \colhead{$\gamma_{k_l}$ [$\mu$m]}
}
\startdata
 3.29 & 0.042  \\
 3.41 & 0.060 \\
 5.70 & 0.20\\
 6.22 & 0.20\\
 6.69 & 0.48\\
 7.60\tablenotemark{a} & 0.34 \\ 
 7.85\tablenotemark{a} & 0.44  \\
 8.33 & 0.50\\
 8.61 & 0.34 \\
 10.68 & 0.22\\
 11.23\tablenotemark{b} & 0.21\\
 11.33\tablenotemark{b} & 0.38 \\
 11.99 &0.54 \\
 12.62 & 0.53 
\enddata
\tablenotetext{a}{the components of the 7.7 $\mu$m complex feature.}
\tablenotetext{b}{the components of the 11.3 $\mu$m complex feature.}
\end{deluxetable}

\begin{deluxetable}{cc}
\tabletypesize{\scriptsize}
%\rotate
\tablecaption{UIR band parameters modeled by Gaussian profiles \label{UIR_g}}
\tablewidth{0pt}
\tablehead{
 \colhead{$\lambda_{k_g}$ [$\mu$m]} & \colhead{$\gamma_{k_g}$ [$\mu$m]}
}
\startdata
 3.46 & 0.034  \\
 3.51 & 0.034 \\
 3.56 & 0.034
\enddata
\end{deluxetable}

\begin{deluxetable}{lcc}
\tabletypesize{\scriptsize}
%\rotate
\tablecaption{Emission line parameters modeled by Gaussian profiles \label{line_g}}
\tablewidth{0pt}
\tablehead{
\colhead{Line} & \colhead{$\lambda_{k_g}$ [$\mu$m]} & \colhead{$\gamma_{k_g}$ [$\mu$m]}
}
\startdata
Br$\beta$ & 2.63 & 0.034  \\
Br$\alpha$ & 4.05 & 0.034  \\
Pf$\gamma$ & 3.74 & 0.034  \\
Pf$\beta$ & 4.65 & 0.034  \\
$[$Ar II$]$ & 6.98 & 0.114 \\
$[$Ar III$]$ & 8.99 & 0.200\\
$[$S IV$]$ & 10.51 & 0.200  \\
$[$Ne II$]$ & 12.81 &0.200 
\enddata
\end{deluxetable}

\begin{deluxetable}{lcccccc}
\tabletypesize{\scriptsize}
%\rotate
\tablecaption{Observed and corrected intensity of the UIR bands\tablenotemark{a}\label{UIR_str}}
\tablewidth{0pt}
\tablehead{
\colhead{Position ID} & \colhead{$I_{\rm 3.3\mu m}$} & \colhead{$I_{\rm 6.2\mu m}$}& \colhead{$I_{\rm 7.7\mu m}$}& \colhead{$I_{\rm 8.6\mu m}$}& \colhead{$I_{\rm 11.3\mu m}$}& \colhead{$I_{\rm 12.6\mu m}$}
}
\startdata
 Position 1   & 0.660$\pm$0.042& 2.12$\pm$0.19 & 6.31$\pm$0.32 & 1.39$\pm$0.14&   3.44$\pm$0.28& 1.11$\pm$0.39 \\
                    & 0.738$\pm$0.046& 2.22$\pm$0.20 & 6.70$\pm$0.34 & 1.62$\pm$0.16&   3.97$\pm$0.32& 1.20$\pm$0.43  \\
 Position 2   & 0.509$\pm$0.063& 1.95$\pm$0.17 & 4.16$\pm$0.27&0.844$\pm$0.120& 2.80$\pm$0.24& 1.04$\pm$0.35 \\
                    & 0.531$\pm$0.065& 1.98$\pm$0.18 & 4.25$\pm$0.28&0.896$\pm$0.130& 2.96$\pm$0.26& 1.08$\pm$0.36 \\
 Position 3   & 0.980$\pm$0.041& 3.40$\pm$0.22 & 9.92$\pm$0.34 & 1.98$\pm$0.16&   4.89$\pm$0.32& 2.95$\pm$0.45  \\
                    & 1.09 $\pm$0.04  & 3.56$\pm$0.23 & 10.5$\pm$0.3   & 2.32$\pm$0.18&   5.66$\pm$0.37& 3.21$\pm$0.49 \\
 Position 4   & 0.693$\pm$0.048& 2.87$\pm$0.15 & 7.15$\pm$0.23 & 1.75$\pm$0.10&   3.27$\pm$0.21& 1.41$\pm$0.30 \\
                    & 0.704$\pm$0.048& 2.89$\pm$0.15 & 7.21$\pm$0.24 & 1.79$\pm$0.10&   3.33$\pm$0.22& 1.43$\pm$0.31 \\
 Position 5   & 0.357$\pm$0.043& 1.32$\pm$0.14 & 3.81$\pm$0.25 & 1.01$\pm$0.11&   1.68$\pm$0.24&- \\
                    & 0.370$\pm$0.049& 1.35$\pm$0.14 & 3.89$\pm$0.26 & 1.08$\pm$0.12&   1.79$\pm$0.26&- \\
 Position 6   & 0.858$\pm$0.035& 3.11$\pm$0.15 & 8.36$\pm$0.25 & 1.79$\pm$0.11&   2.99$\pm$0.22& 0.99$\pm$0.32 \\
                    & 0.891$\pm$0.036& 3.15$\pm$0.15 & 8.53$\pm$0.25 & 1.88$\pm$0.11&   3.13$\pm$0.23& 1.02$\pm$0.33 \\
 Position 7   & 4.85 $\pm$0.04  & 23.1$\pm$0.2   & 63.7$\pm$0.3   & 13.0$\pm$0.1  &   25.3$\pm$0.3  & 11.92$\pm$0.45 \\
                    & 4.81 $\pm$0.05  & 23.4$\pm$0.2   & 64.2$\pm$0.3   & 13.6$\pm$0.1  &   26.4$\pm$0.3  & 12.21$\pm$0.46 \\
 Position 8-1& 4.07 $\pm$0.07  & 14.5$\pm$0.2  & 52.3$\pm$0.5    & 13.4$\pm$0.2 &   28.3$\pm$0.5  & 15.04$\pm$0.74\\
                    & 4.44 $\pm$0.09  & 15.3$\pm$0.2  & 55.4$\pm$0.5    & 16.0$\pm$0.2 &   33.2$\pm$0.6  & 16.52$\pm$0.82 \\
 Position 8-2& 1.13 $\pm$0.10  & 4.76$\pm$0.39 & 17.8$\pm$0.6   & 4.79$\pm$0.27&   7.01$\pm$0.57& 3.05$\pm$0.83  \\
                    & 1.16 $\pm$0.11  & 5.01$\pm$0.41 & 18.6$\pm$0.6   & 5.70$\pm$0.33&   8.24$\pm$0.67& 3.35$\pm$0.91 
\enddata
\tablecomments{The UIR bands detected with more than 2\,$\sigma$ are listed. 
For each position, the upper row indicates the original band intensity 
and the lower row shows the intensity for which the extinction and the contribution from the hydrogen recombination lines are corrected (see text).}
\tablenotetext{a}{In units of 10$^{-18}$W m$^{-2}$ arcsec$^{-2}$.}
\end{deluxetable}

\begin{deluxetable}{lcccccccc}
\tabletypesize{\scriptsize}
%\rotate
\tablecaption{Observed and corrected intensity of the emission lines\tablenotemark{a}\label{el_str}}
\tablewidth{0pt}
\tablehead{
\colhead{Position ID}& \colhead{$I_{\rm Br\beta2.63}$} & \colhead{$I_{\rm Br\alpha4.05}$} &  \colhead{$I_{\rm Pf\gamma3.74}$} & \colhead{$I_{\rm Pf\beta4.65}$} &\colhead{$I_{\rm[Ar II]6.98}$}& \colhead{$I_{\rm[Ar III]8.99}$}& \colhead{$I_{\rm[S IV]10.51}$}& \colhead{$I_{\rm[Ne II]12.81}$}
}
\startdata
Position 1	&	-	&	-	&	-	&	-	&	-	&	-	&	-	&	-\\
Position 2	&	-	&	-	&	-	&	-	&	-	&	-	&	-	&	-\\
Position 3	&	-	&	-	&	-	&	-	&	-	&	-	&	-	&	-\\
Position 4	&	-	&	-	&	-	&	-	&	-	&	-	&	-	&	-\\
Position 5	&	-	&	0.0599$\pm$0.0159	&	-	&	-	&	-	&	-	&	-	&	-\\
                 &	-	&	0.0610$\pm$0.0160	&	-	&	-	&	-	&	-	&	-	&	-\\
Position 6	&	-	&	-	&	-	&	-	&	-	&	-	&	-	&	-\\
Position 7    &  1.02$\pm$0.03   & 2.04$\pm$0.01 & 0.261$\pm$0.017 & 0.457$\pm$0.022&     1.65$\pm$0.07& 1.48$\pm$0.05&   1.56$\pm$0.05  & 3.94 $\pm$0.12 \\
                    &  1.07$\pm$0.03   & 2.08$\pm$0.01 & 0.267$\pm$0.018 & 0.464$\pm$0.023&     1.67$\pm$0.08& 1.57$\pm$0.05&   1.65$\pm$0.05  & 4.03 $\pm$0.12 \\
 Position 8-1& 0.832$\pm$0.053& 1.70$\pm$0.02 & 0.285$\pm$0.034 & 0.279$\pm$0.029 & 0.780$\pm$0.105& 1.99$\pm$0.08& 0.948$\pm$0.082 & 3.63$\pm$0.20 \\
                    & 0.996$\pm$0.064& 1.85$\pm$0.03 & 0.315$\pm$0.037 & 0.297$\pm$0.030 & 0.812$\pm$0.109& 2.51$\pm$0.11&   1.18$\pm$0.10   & 3.97$\pm$0.22 \\
Position 8-2& 0.631$\pm$0.067 & 1.27$\pm$0.03 & 0.141$\pm$0.036 & 0.262$\pm$0.042& 0.482$\pm$0.145& 1.39$\pm$0.09&   1.26$\pm$0.09  & 1.83 $\pm$ 0.23 \\
                   & 0.755$\pm$0.080 & 1.38$\pm$0.03 & 0.156$\pm$0.040 & 0.279$\pm$0.045& 0.503$\pm$0.151& 1.75$\pm$0.12&   1.57$\pm$0.11  & 2.00 $\pm$ 0.25 
\enddata
\tablecomments{Emission lines detected with more than 2\,$\sigma$ are listed. 
For each position, the upper row indicates the original band intensity and the lower row shows the extinction-corrected intensity (see text).}
\tablenotetext{a}{In units of 10$^{-18}$W m$^{-2}$ arcsec$^{-2}$.}
\end{deluxetable}

\begin{deluxetable}{lcccccc}
\tabletypesize{\scriptsize}
%\rotate
\tablecaption{The UIR band intensity ratios  \label{UIR_ratio_e_c}}
\tablewidth{0pt}
\tablehead{
\colhead{Position ID} & \colhead{$I_{\rm 3.3\,\mu m}$/$I_{\rm 11.3\,\mu m}$} & \colhead{$I_{\rm 6.2\,\mu m}$/$I_{\rm 11.3\,\mu m}$}& \colhead{$I_{\rm 7.7\,\mu m}$/$I_{\rm 11.3\,\mu m}$}& \colhead{$I_{\rm 8.6\,\mu m}$/$I_{\rm 11.3\,\mu m}$}& \colhead{$I_{\rm 12.6\,\mu m}$/$I_{\rm 11.3\,\mu m}$}& \colhead{$I_{\rm 6.2\,\mu m}$/$I_{\rm 7.7\,\mu m}$}
}
\startdata
 Position 1    & 0.185$\pm$0.019 & 0.559$\pm$0.069 & 1.68$\pm$0.16 & 0.409$\pm$0.052 & 0.304$\pm$0.111& 0.331$\pm$0.035 \\
 Position 2    & 0.179$\pm$0.027 & 0.671$\pm$0.085 & 1.43$\pm$0.16 & 0.303$\pm$0.051 & 0.365$\pm$0.129& 0.466$\pm$0.053 \\
 Position 3    & 0.193$\pm$0.015 & 0.629$\pm$0.058 & 1.86$\pm$0.13 & 0.411$\pm$0.042 & 0.567$\pm$0.095& 0.337$\pm$0.025 \\
 Position 4    & 0.211$\pm$0.020 & 0.867$\pm$0.073 & 2.16$\pm$0.15 & 0.537$\pm$0.048 & 0.428$\pm$0.097& 0.401$\pm$0.025  \\
 Position 5    & 0.205$\pm$0.040 & 0.752$\pm$0.136 & 2.16$\pm$0.38 & 0.601$\pm$0.112 &-                              & 0.347$\pm$0.044  \\
 Position 6    & 0.284$\pm$0.024 & 1.01  $\pm$0.09   & 2.71$\pm$0.21 & 0.591$\pm$0.057 & 0.326$\pm$0.108& 0.370$\pm$0.021  \\
 Position 7    & 0.182$\pm$0.003 & 0.887$\pm$0.013 & 2.42$\pm$0.03 & 0.514$\pm$0.009 & 0.461$\pm$0.018& 0.365$\pm$0.003  \\
 Position 8-1 & 0.133$\pm$0.003 & 0.461$\pm$0.096 & 1.66$\pm$0.33 & 0.481$\pm$0.012 & 0.496$\pm$0.026& 0.276$\pm$0.006  \\
 Position 8-2 & 0.141$\pm$0.018 & 0.608$\pm$0.069 & 2.26$\pm$0.20 & 0.692$\pm$0.068 & 0.406$\pm$0.115& 0.268$\pm$0.023 
\enddata
\tablecomments{The band intensity ratio for which the extinction and the contribution from the hydrogen recombination lines are corrected. 
}
\end{deluxetable}

\clearpage

%1->3->1:1400318.1
%5->1->2:1400330.1
%4->2->3:1400346.1
%8->4:1402426.1
%3->5:1400324.1
%7->6:1402422.1
%6->7:1400334.1
%2->8:1400320.1

%Position 1 & 1.88988    \\
%Position 2 &  0.72118   \\
%Position 3 & 1.90918    \\
%Position 4 & 0.2576066\\
%Position 5 & 0.826695  \\
%Position 6 & 0.6302435\\
%Position 7 & 0.54175    \\
%Position 8-1 &2.11669 \\
%Position 8-2 & 2.11669 

\begin{figure}
\epsscale{1.}
\plotone{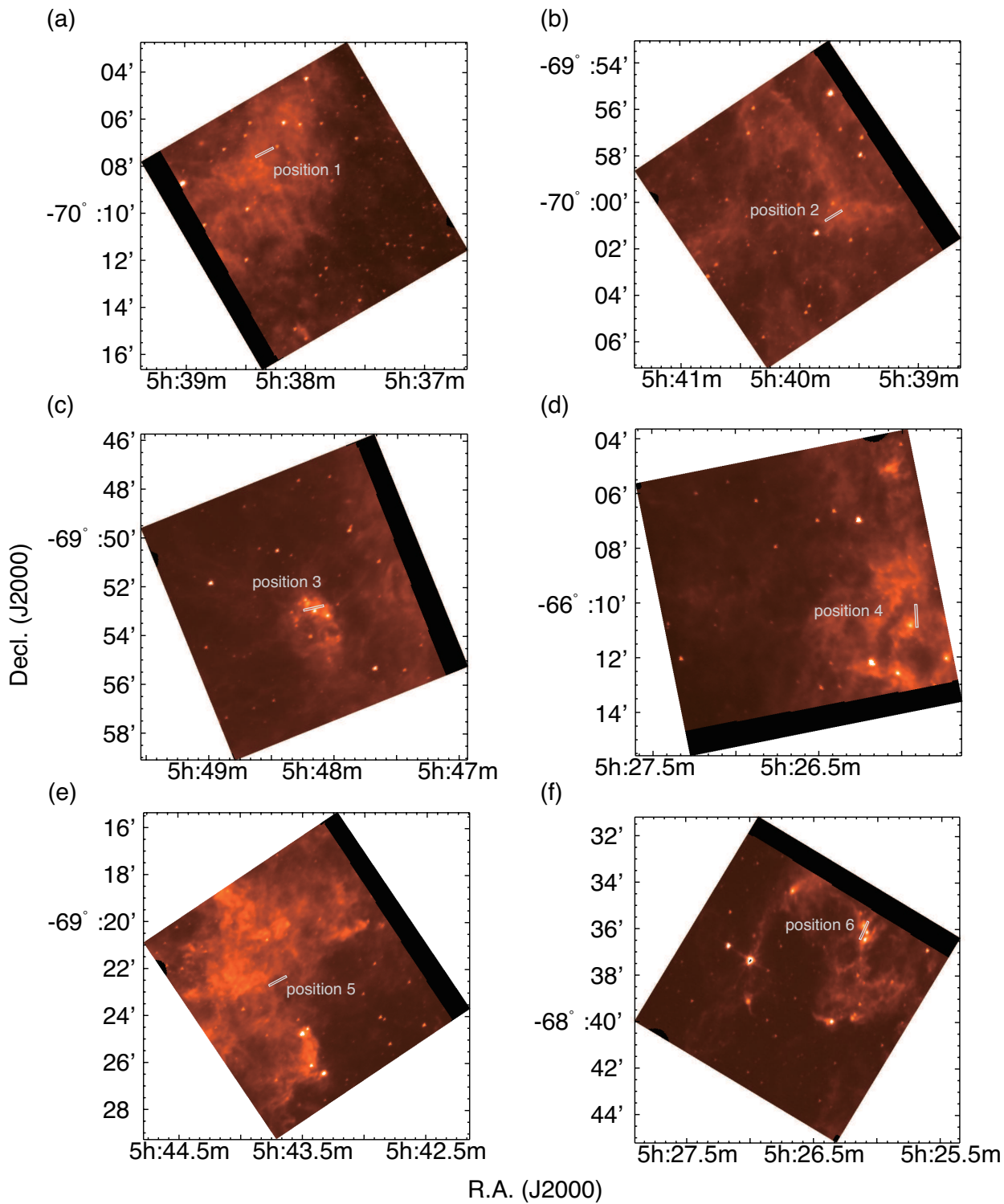}
\caption{The enlarged slit position images of {\bf (a)}Position 1, {\bf (b)}Position 2, {\bf (c)}Position 3, {\bf (d)}Position 4, {\bf (e)}Position 5, {\bf (f)}Position 6, {\bf (g)}Position 7, and {\bf (h)}Position 8 are overlain on the S11 images of the LMC. 
The white boxes indicate the slit of $1\arcmin$ length by $5\arcsec$ width in each observation. 
\label{s11}
}
\end{figure}

\setcounter{figure}{1}
\begin{figure}
\epsscale{1.}
\plotone{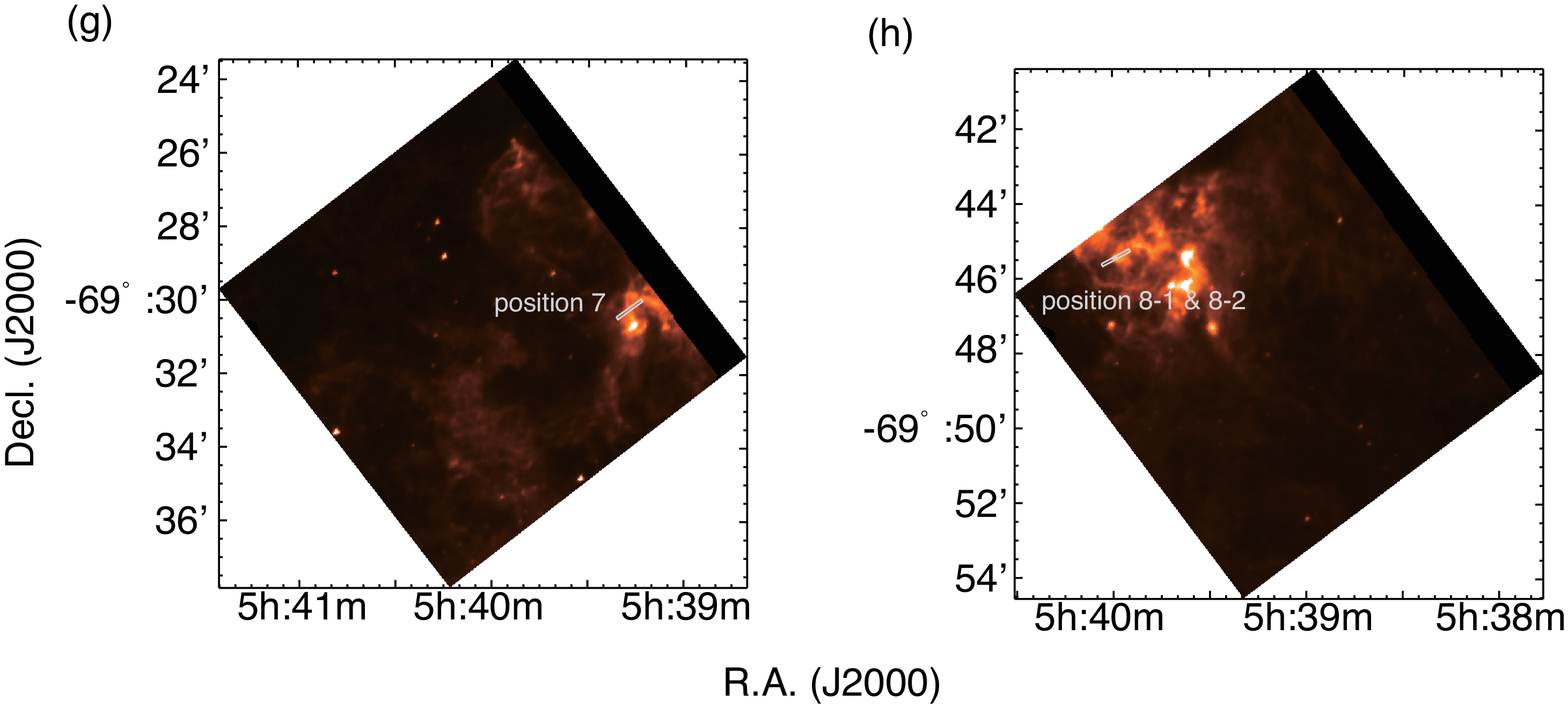}
\caption{Continued.}
\end{figure}

\begin{figure}
\epsscale{.60}
\plotone{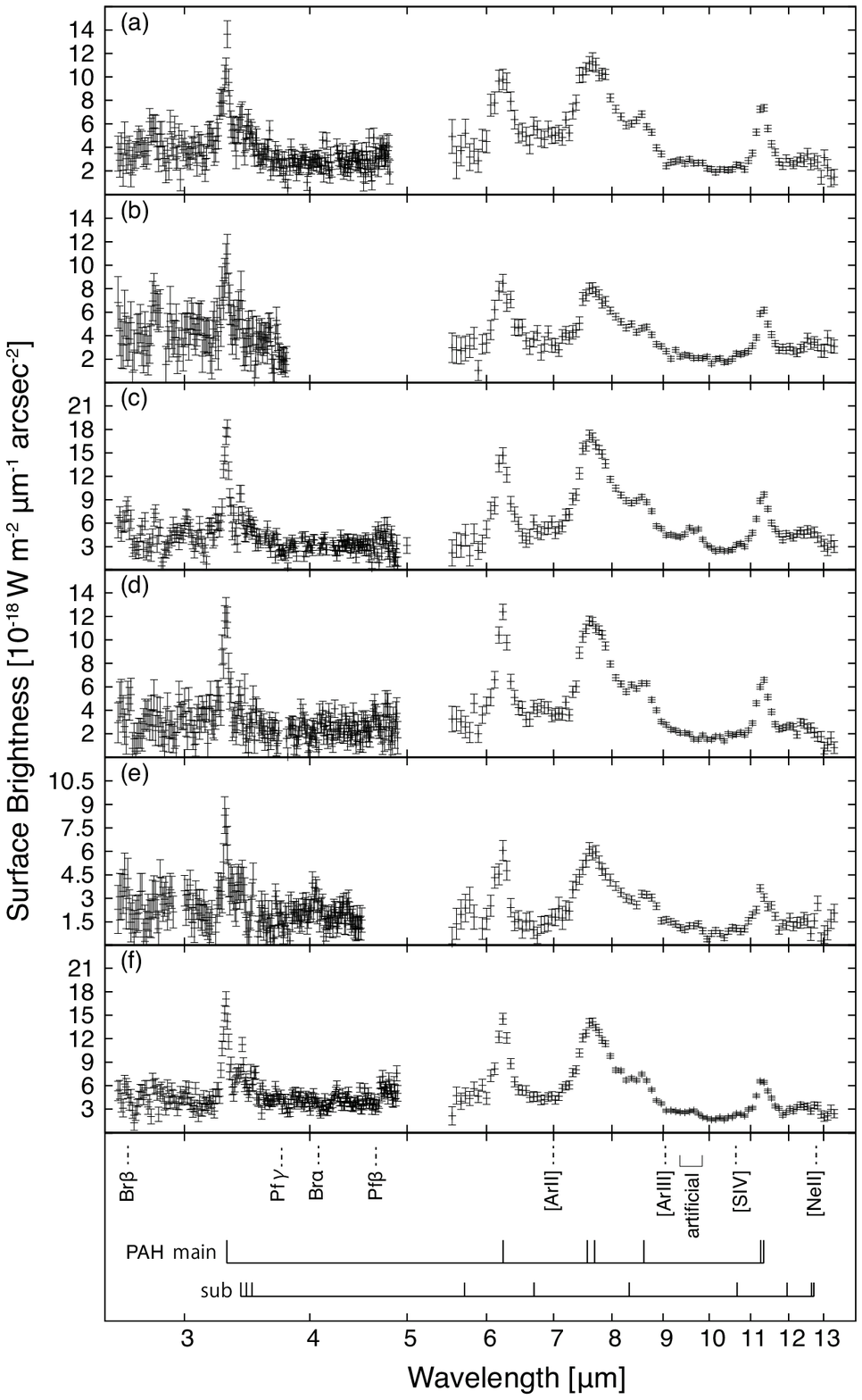}
\caption{Observed spectra of nine regions in the LMC; {\bf (a)} Position 1, {\bf (b)} Position 2, {\bf (c)} Position 3, {\bf (d)} Position 4, {\bf (e)} Position 5, {\bf (f)} Position 6 {\bf (g)} Position 7, {\bf (h)} Position 8-1, and {\bf (i)} Position 8-2. 
The bottom panel indicates the positions of the UIR bands and the emission lines. 
\label{fig_lmc_raw}
}
\end{figure}

\setcounter{figure}{2}
\begin{figure}
\epsscale{.60}
\plotone{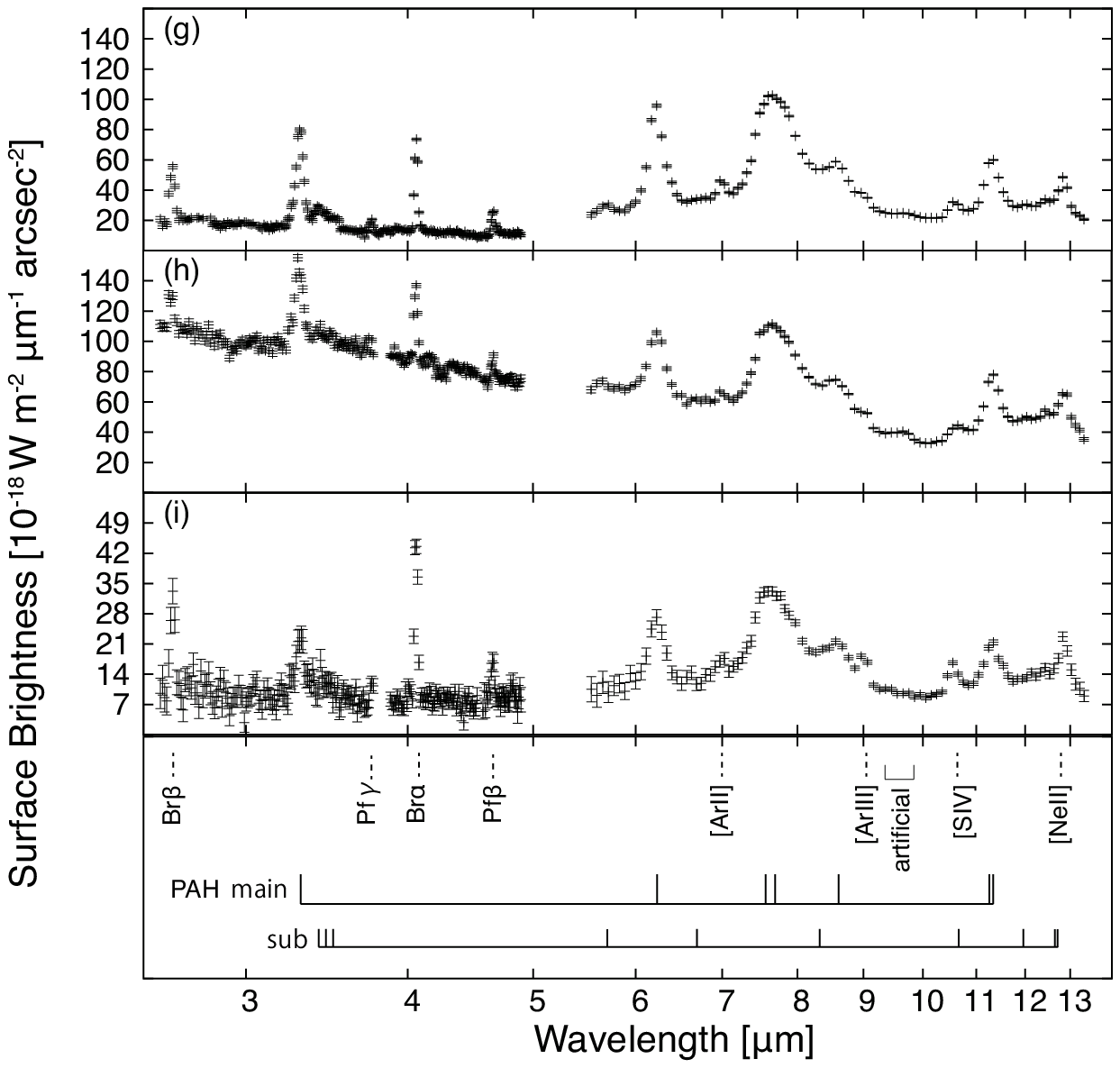}
\caption{Continued.}
\end{figure}

\begin{figure}
\epsscale{.60}
\plotone{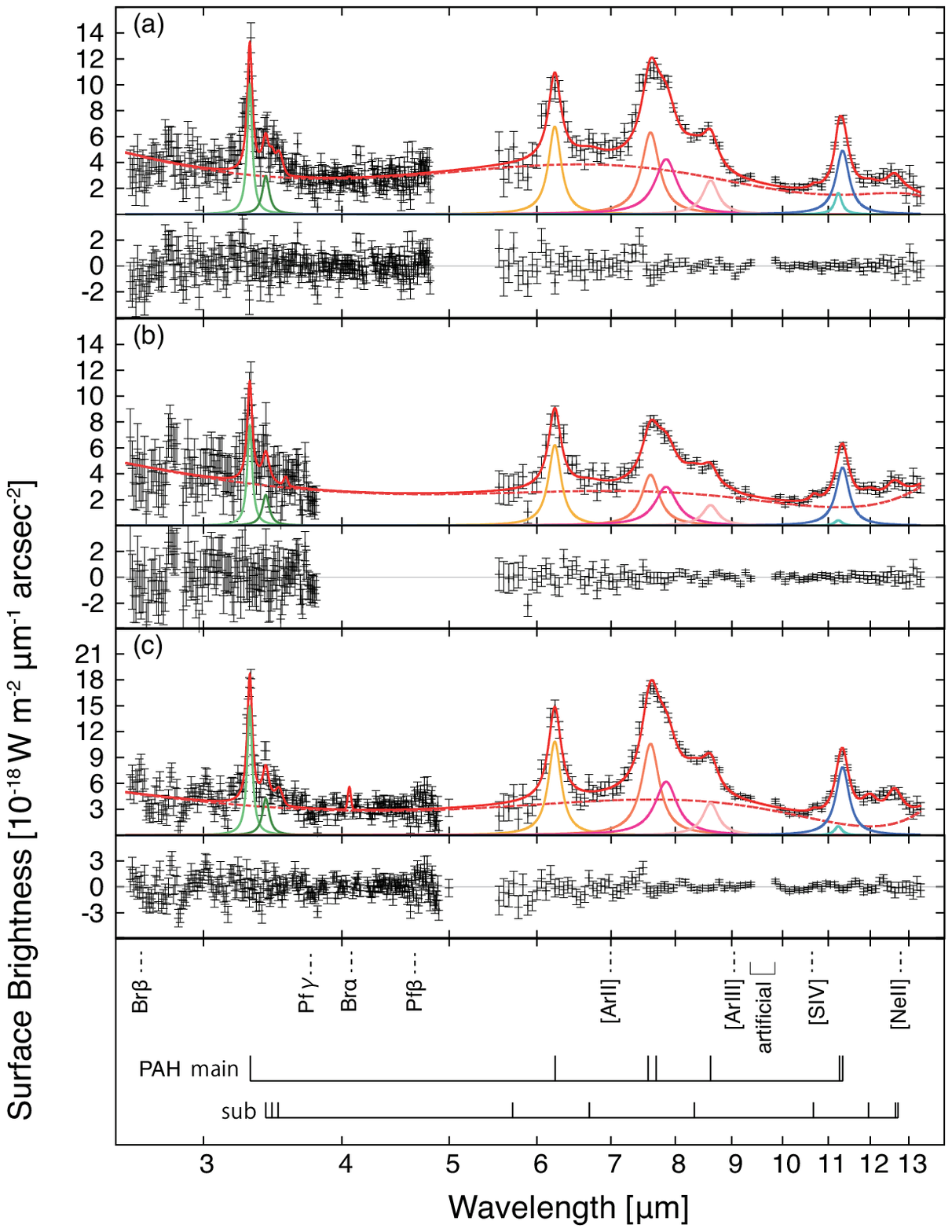}
\caption{Modeled spectra are overlain on the observed spectra of nine regions in the LMC; {\bf (a)} Position 1, {\bf (b)} Position 2, {\bf (c)} Position 3, {\bf (d)} Position 4, {\bf (e)} Position 5, {\bf (f)} Position 6 {\bf (g)} Position 7, {\bf (h)} Position 8-1, and {\bf (i)} Position 8-2. The red solid lines show the best-fit model spectra (see text). The light-red dotted lines indicate the continuum component of a quintic function. The light-green, dark-green, yellow, coral, dark-pink, pink, light-blue and dark-blue lines show the UIR 3.3\,$\mu$m, 3.4\,$\mu$m, 6.2\,$\mu$m, 7.6\,$\mu$m, 7.8\,$\mu$m, 8.6\,$\mu$m, 11.2\,$\mu$m and 11.3\,$\mu$m bands of a Lorentzian function, respectively. We cut 9.4--9.8\,$\mu$m from spectra in Figure \ref{fig_lmc_raw} due to the artifacts (see text). The lower panel shows a residual spectrum at each plot. 
The bottom panel is the same as in Figure \ref{fig_lmc_raw}. 
\label{fig_lmc_spec_fit}
}
\end{figure}

\setcounter{figure}{3}
\begin{figure}
\epsscale{.60}
\plotone{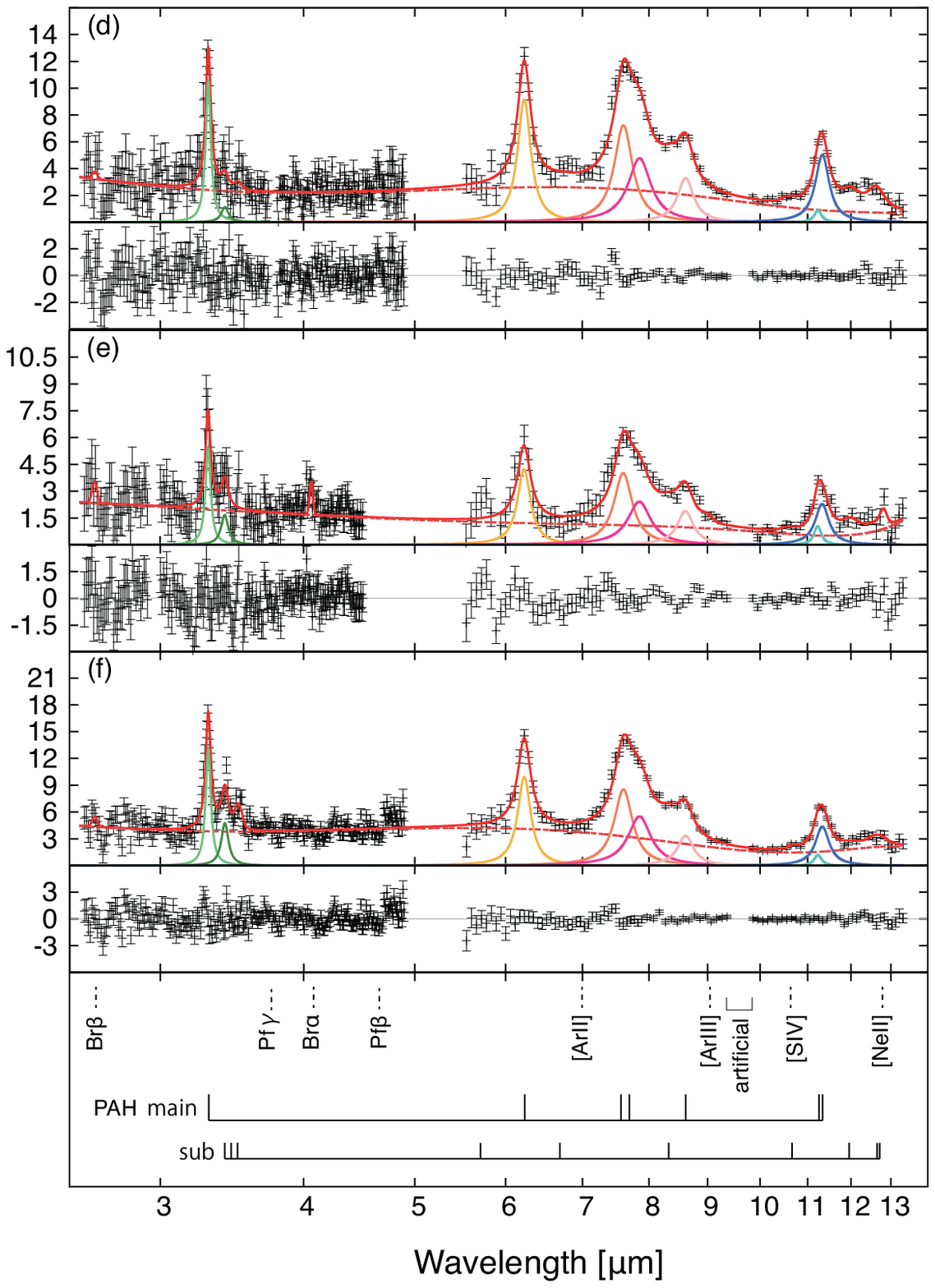}
\caption{Continued.}
\end{figure}

\setcounter{figure}{3}
\begin{figure}
\epsscale{.60}
\plotone{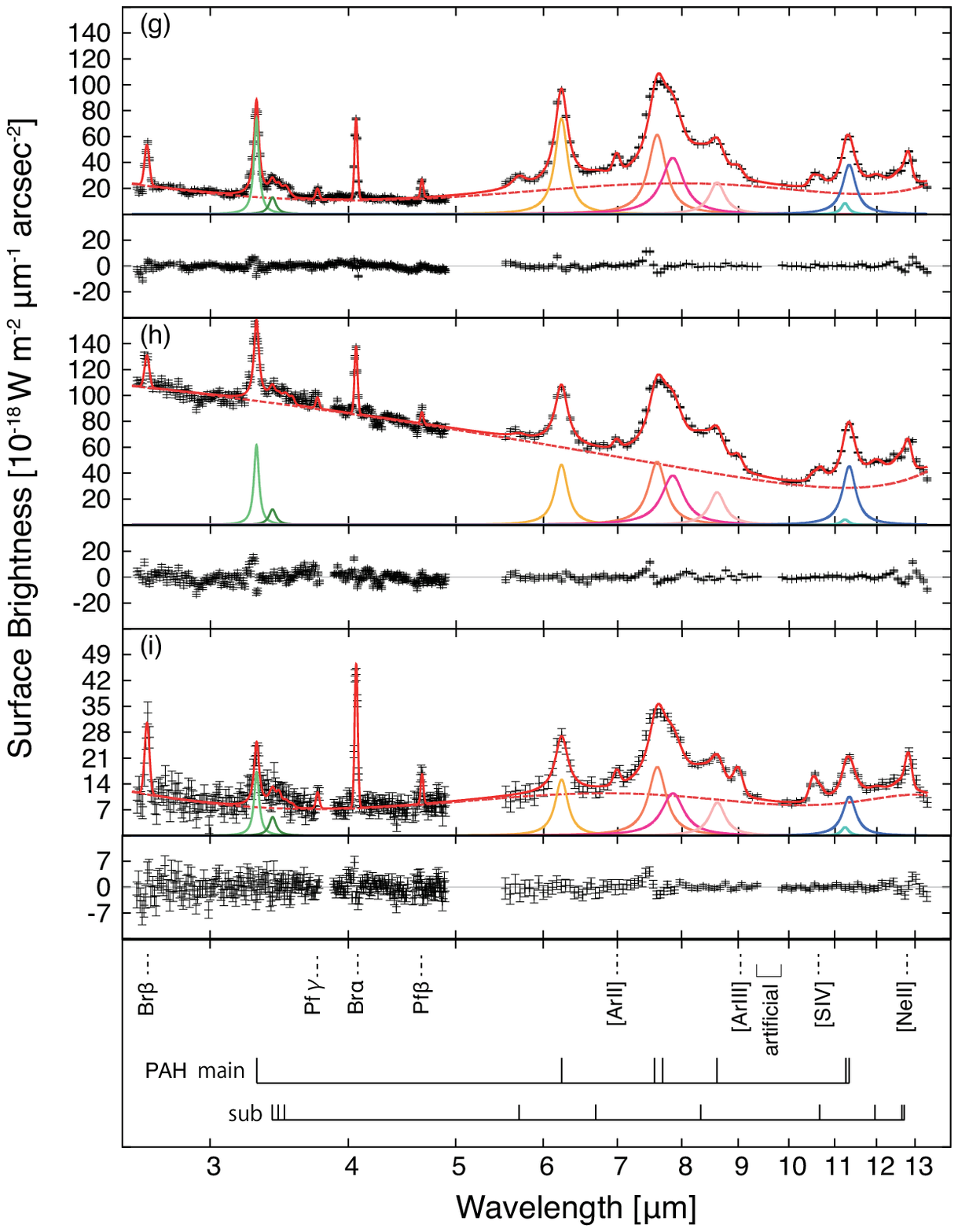}
\caption{Continued.}
\end{figure}

\begin{figure}
\epsscale{0.8}
\plotone{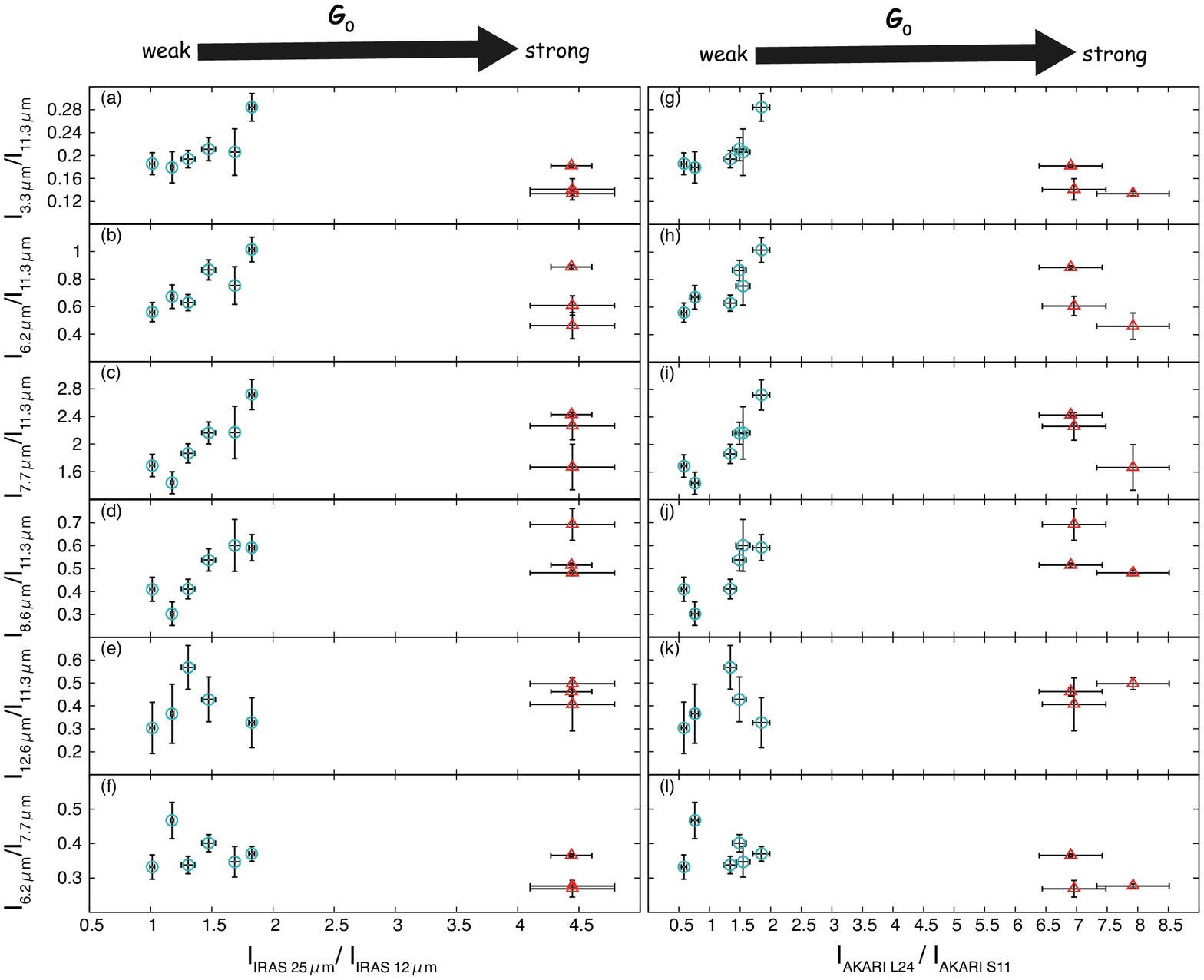}
\caption{ Variation of the UIR band ratios of the 3.3 $\mu$m to the 11.3 $\mu$m band ({\bf (a)} and {\bf (g)}), the 6.2\,$\mu$m to the 11.3\,$\mu$m band ({\bf (b)} and {\bf (h)}), the 7.7\,$\mu$m to the 11.3\,$\mu$m band ({\bf (c)} and {\bf (i)}), the 8.6\,$\mu$m to the 11.3\,$\mu$m band ({\bf (d)} and {\bf (j)}), the 12.6\,$\mu$m to the 11.3\,$\mu$m band ({\bf (e)} and {\bf (k)}), and the 6.2\,$\mu$m to the 7.7\,$\mu$m band ({\bf (f)} and {\bf (l)}) against the {\it IRAS} color of $I_{\rm 25\,\mu m}$/$I_{\rm 12\,\mu m}$ and the {\it AKARI} color of $I_{\rm L24}$/$I_{\rm S11}$. 
The turquoise open circles and the red open triangles with the black errorbars indicate the UIR band ratios and the AKARI or IRAS colors of Groups A and B, respectively. 
The UIR band ratios are corrected for the extinction and the contribution from the hydrogen recombination lines. 
\label{UIRratio_IRAS}}
\end{figure}

\begin{figure}
\epsscale{0.8}
\plotone{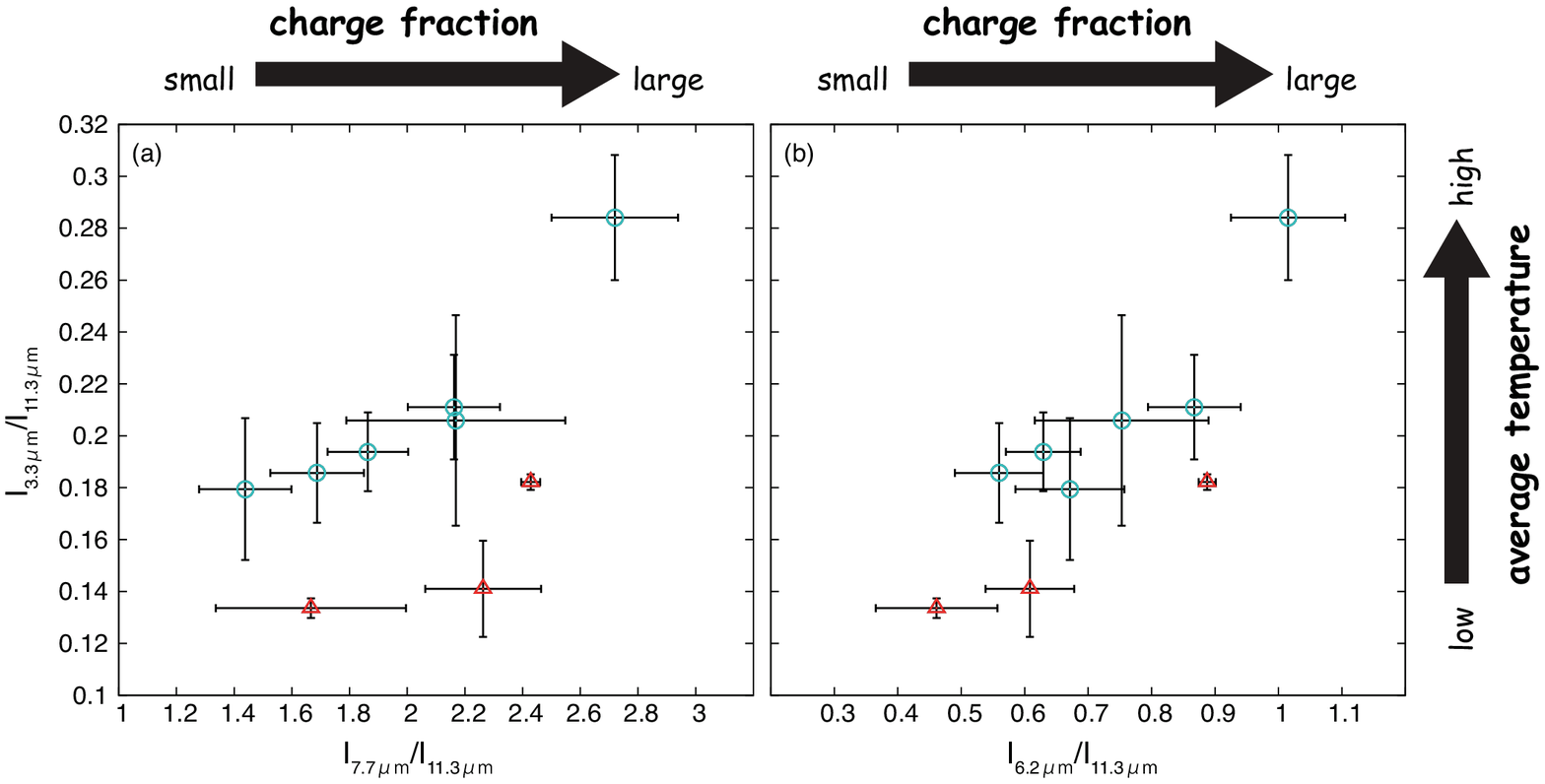}
\caption{Variation of the UIR band ratios of {\bf (a)} the 7.7\,$\mu$m to the 11.3\,$\mu$m band and {\bf (b)} the 6.2\,$\mu$m band the 11.3\,$\mu$m band against the UIR band ratio of the 3.3\,$\mu$m to the 11.3\,$\mu$m band. 
The turquoise open circles and the red open triangles with the black errorbars indicate the UIR band ratios of Groups A and B, respectively. 
The UIR band ratios are corrected for the extinction and the contribution from the hydrogen recombination lines. 
\label{UIRratio_UIRratio}
}
\end{figure}

\begin{figure}
\epsscale{0.8}
\plotone{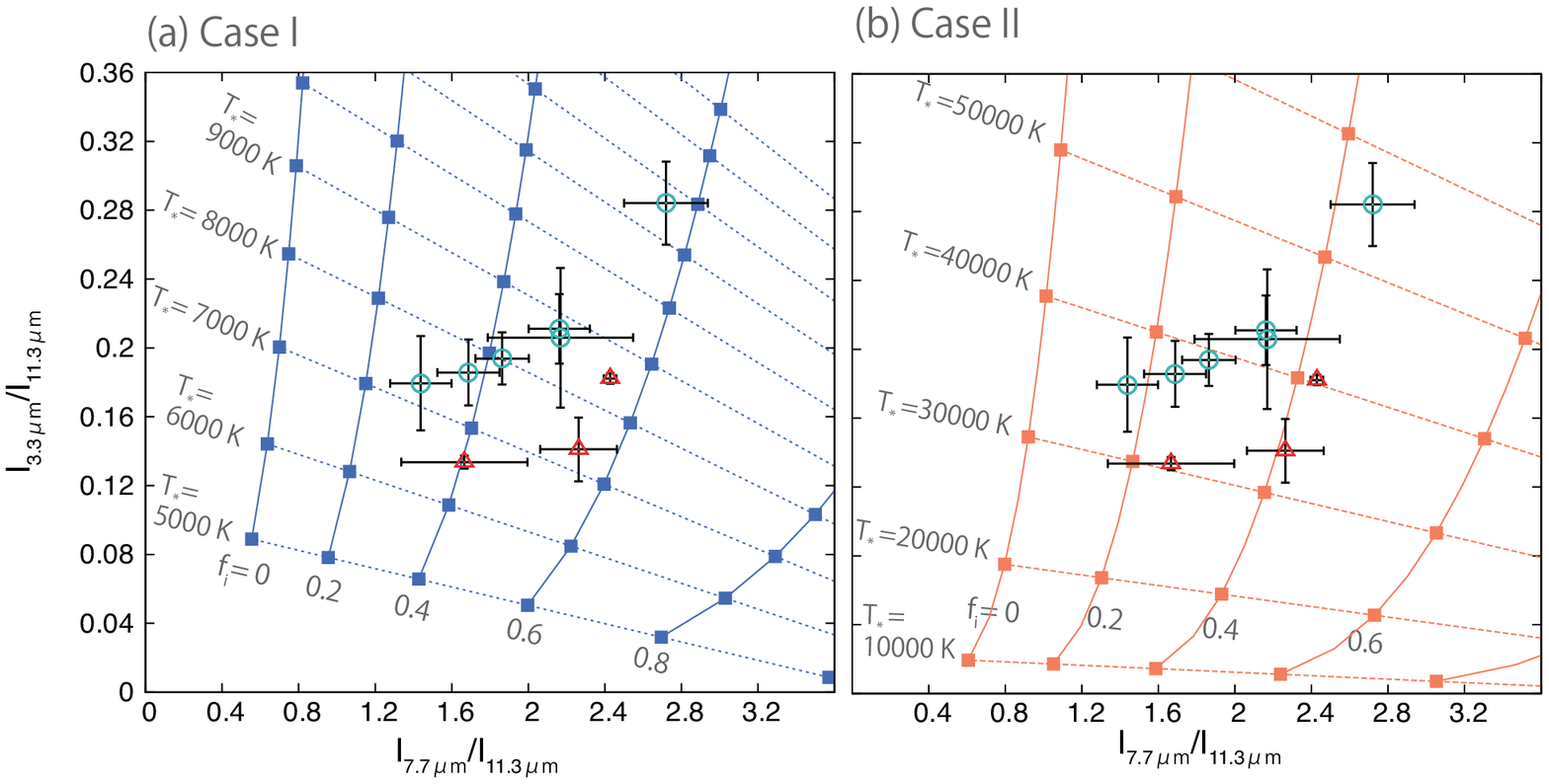}
\caption{Comparison of the observed UIR band ratios with the models. 
The model grids are calculated for a mixture of neutral and ionized PAHs with a power-law size distribution exposed to a blackbody with different temperatures $T_*$ and various ionized fractions of PAHs $f_{\rm i}$. Case I {\bf (a)} is calculated with ($n_{\rm C}^{\rm min}$, $n_{\rm C}^{\rm max}$)=(20,4000) and Case II {\bf (b)} with ($n_{\rm C}^\mathrm{min}$, $n_{\rm C}^\mathrm{max}$) = (100,4000). 
See Appendix for details of the model calculation. 
The symbols of the observed points are the same as in Figure \ref{UIRratio_UIRratio}. 
\label{model_ratio}
}
\end{figure}

\end{document}